\documentclass{aa}  
\usepackage[utf8]{inputenc}
\usepackage[T1]{fontenc}
\usepackage{color}
\usepackage{multirow}
\usepackage{afterpage}
\usepackage{amsmath}
\usepackage{multicol}

\usepackage{graphicx}
\usepackage{textcomp}
\usepackage{ragged2e}

\usepackage{txfonts}
\usepackage{xcolor}

\usepackage{placeins}
\usepackage{nicefrac}
\usepackage{gensymb}
\usepackage{wasysym}
\usepackage{natbib,twoopt}
\usepackage[breaklinks=true]{hyperref} %% to avoid \citeads line fills
\bibpunct{(}{)}{;}{a}{}{,} %% natbib format for A&A and ApJ
\makeatletter
\newcommandtwoopt{\citeads}[3][][]{\href{http://adsabs.harvard.edu/abs/#3}%
{\def\hyper@linkstart##1##2{}%
\let\hyper@linkend\@empty\citealp[#1][#2]{#3}}}
\newcommandtwoopt{\citepads}[3][][]{\href{http://adsabs.harvard.edu/abs/#3}%
{\def\hyper@linkstart##1##2{}%
\let\hyper@linkend\@empty\citep[#1][#2]{#3}}}
\newcommandtwoopt{\citetads}[3][][]{\href{http://adsabs.harvard.edu/abs/#3}%
{\def\hyper@linkstart##1##2{}%
\let\hyper@linkend\@empty\citet[#1][#2]{#3}}}
\newcommandtwoopt{\citeyearads}[3][][]%
{\href{http://adsabs.harvard.edu/abs/#3}
{\def\hyper@linkstart##1##2{}%
\let\hyper@linkend\@empty\citeyear[#1][#2]{#3}}}
\makeatother

\hypersetup{
    colorlinks,
   linkcolor={red!60!black},
   citecolor={blue!60!black},
    urlcolor={blue!60!black}
}

\usepackage{footmisc}

\begin{document} 
	
\title{Simulating the environment around planet-hosting stars}
\subtitle{II. Stellar winds and inner astrospheres}

\author{J. D. Alvarado-G\'omez\inst{1,2}, G. A. J. Hussain\inst{1,3}, O. Cohen\inst{4}, J. J. Drake\inst{4}, C. Garraffo\inst{4}, J. Grunhut\inst{1} 
         	%\inst{1} 
	\and	
	 T. I. Gombosi\inst{5}%\fnmsep\thanks{Just to show the usage
          %of the elements in the author field}
          }
          \institute{\inst{1} European Southern Observatory,
              Karl-Schwarzschild-Str. 2, 85748 Garching bei M\"unchen, Germany\\
              \email{jalvarad@eso.org} \\
              \inst{2} Universit\"ats-Sternwarte, Ludwig-Maximilians-Universit\"at M\"unchen, Scheinerstr.~1, 81679 M\"unchen, Germany \\
               \inst{3} Institut de Recherche en Astrophysique et Plan\'etologie, Universit\'e de Toulouse, UPS-OMP, F-31400 Toulouse, France \\
              \inst{4} Harvard-Smithsonian Center for Astrophysics, 60 Garden Street, Cambridge, MA 02138, USA\\
              \inst{5} Center for Space Environment Modeling, University of Michigan, 2455 Hayward St., Ann Arbor, MI 48109, USA	
             }
             
   \date{Received -----; accepted -----}

% \abstract{}{}{}{}{} 
% 5 {} token are mandatory
 
\abstract{We present the results of a comprehensive numerical simulation of the environment around three exoplanet-host stars (HD\,1237, HD\,22049, and HD\,147513). Our simulations consider one of the latest models currently used for space weather studies in the Heliosphere, with turbulent Alfv\'en wave dissipation as the source of coronal heating and stellar wind acceleration. Large-scale magnetic field maps, recovered with two implementations of the tomographic technique of Zeeman-Doppler imaging, serve to drive steady-state solutions in each system. This paper contains the description of the stellar wind and inner astrosphere, while the coronal structure was previously discussed in \citetads{2016A&A...588A..28A}. The analysis includes the magneto-hydrodynamical properties of the stellar wind, the associated mass and angular momentum loss rates, as well as the topology of the astrospheric current sheet in each system. A systematic comparison among the considered cases is performed, including two reference solar simulations covering activity minimum and maximum. For HD\,1237, we investigate the interactions between the structure of the developed stellar wind, and a possible magnetosphere around the Jupiter-mass planet in this system. We find that the process of particle injection into the planetary atmosphere is dominated by the density distribution rather than velocity profile of the stellar wind. In this context, we predict a maximum exoplanetary radio emission of 12 mJy at 40 MHz in this system, assuming the crossing of a high-density streamer during periastron passage. Furthermore, in combination with the analysis performed in \citetads{2016A&A...588A..28A}, we obtain for the first time a fully simulated mass loss-activity relation, which is compared and discussed in the context of the relation based on astrospheric detections proposed by \citetads{2005ApJ...628L.143W}. Finally, we provide a characterisation of the global 3D properties of the stellar wind of these systems, at the inner edges of their habitable zones.}   


\titlerunning{Stellar winds and inner astrospheres}
\authorrunning{Alvarado-G\'omez et al.}
\maketitle
%
%________________________________________________________________

\section{Introduction}\label{sec_intro}

\noindent As well as driving stellar activity cycles, magnetic fields strongly influence different aspects of the stellar structure and evolution. It is known that they play a major role in the coronal heating processes in the Sun and other late type stars (\citeads{2015RSPTA.37340269D}; \citeads{2015RSPTA.37340259T}), as well as in the generation of persistent stellar winds and astrospheres (see \citeads{2004LRSP....1....2W}). These stellar winds are crucial to understand the evolution of rotation and magnetic activity in cool stars on the early main sequence. G to K type stars tend to rotate rapidly on the Zero Age Main Sequence (ZAMS); braking torques exerted by winds cause them to spin down, losing most of their angular momentum within the first 500 Myr (\citeads{2016A&A...587A.105A}; \citeads{2010ApJ...721..675B}). Strong winds from the young Sun have been used to explain both the stripping of the Martian atmosphere (\citeads{2013oepa.book.....L}; \citeads{2009AsBio...9...55T}), and address the ``faint young Sun paradox''. This paradox is that terrestrial geological records indicate that water existed in liquid form very early in the history of Earth and Mars, despite the young Sun having only 70\% of its current luminosity. Solar wind sputtering is a leading candidate to explain the loss of Mars' once-thick atmosphere, because Mars is not protected by a strong magnetosphere, unlike Earth (see \citeads{2007SSRv..129..245L}). However, recent work presented by \citetads{2014ApJ...781L..33W} argues that while the young Sun was more magnetically active, it does not necessarily follow that it would have hosted stronger winds. 

This last result comes from the close relation between the winds in Sun-like stars and their surrounding astrospheres. In the case of the Sun, the solar wind creates a comet-like bubble (the heliosphere) that extends far past the orbits of the planets, and interacts with the local interstellar medium (LISM)\footnote[2]{This classical shape of the heliosphere has been recently revisited in various observational and numerical works, pointing towards a far more complex description including magnetized jets (see \citeads{2013ApJ...771...77M};  \citeads{2015ApJ...800L..28O}; \citeads{2015ApJ...808L..44D}).}. The heliosphere is populated by hot hydrogen atoms created through charge exchange between the ionized gas in the solar wind and the cold LISM hydrogen. Hot hydrogen builds up particularly in the region between the termination shock and the heliopause. This is the region which the Voyager mission may recently have crossed (\citeads{2013Sci...341.1489G}), although this is still a matter of debate (see \citeads{2014ApJ...789...41F}; \citeads{2015ApJ...806L..27G}). This hydrogen wall is detected as extra H\,I Lyman-$\alpha$ absorption in the UV spectra of cool stars. Stronger winds result in a larger astrosphere and increased absorption (\citeads{2014ASTRP...1...43L}). By measuring the column densities and velocities of this extra absorption it is possible to derive the only available observationally-driven estimates of mass loss rates in cool stars (see \citeads{2015ASSL..411...19W}). However, these estimates strongly depend on the assumed characteristics and topology of the LISM (see \citeads{2014ASTRP...1...43L}), for which there is still no complete agreement in the literature (e.g., \citeads{2009ApJ...696.1517K}; \citeads{2014A&A...567A..58G}; \citeads{2015ApJ...812..125R}).

On the theoretical and modelling side, recent studies have provided different frameworks for the stellar wind origin, behaviour, and influence in the angular momentum evolution of late-type stars. Among the 1D and 2D models, a non-comprehensive list includes semi-empirical approaches for thermally-driven winds, within a hydro- (e.g. Johnstone et al. \citeyearads{2015A&A...577A..27J}, \citeyearads{2015A&A...577A..28J}) or magneto-hydrodynamic (MHD) regime (e.g. Matt et al. \citeyearads{2008ApJ...678.1109M}, \citeyearads{2012ApJ...754L..26M}; R\'eville et al. \citeyearads{2015ApJ...798..116R}, \citeyearads{2015ApJ...814...99R}), physically-motivated descriptions involving scaling relations for the stellar magnetic fields, rotation periods, convective properties, and X-ray fluxes (e.g. \citeads{2012ApJ...746...43R}; \citeads{2016MNRAS.458.1548B}), and semi-analytic and numerical formulations based on Alfv\'en wave MHD turbulence (e.g. \citeads{2011ApJ...741...54C}; \citeads{2013PASJ...65...98S}). While providing reasonable agreement in the rotational evolution of late-type stars at different stages (e.g. \citeads{2013A&A...556A..36G}, \citeads{2015ApJ...799L..23M}), such approaches are very generic and cannot capture the specifics of the stellar wind of a given system. The same is true for the complex interplay between the magnetic field topology, coronal structure, and the stellar wind. These elements are fundamental for a better understanding of the environment around planet-hosting stars, including the relative influence of the wind and the high-energy emission on the exoplanetary conditions and habitability (see \citeads{2003ApJ...598L.121L}; \citeads{2013oepa.book.....L}; \citeads{2014ApJ...795..132S}; \citeads{2014RSPTA.37230084F}). Such detailed descriptions are crucial for the current and future perspectives in the area of exoplanetary characterisation from the ground and space (see \citeads{2014Natur.513..358P}; \citeads{2014Natur.513..353H}). 

In this context, we presented in \citetads{2016A&A...588A..28A} the initial results of a detailed 3D numerical study aimed at simulating the environment around planet-hosting stars. This previous article described the developed coronal structure and high-energy environment on three exoplanet-hosts, namely HD\,22049 (K2V), HD\,1237 (G8V), and HD\,147513 (G5V). The basic stellar and planetary (orbital) properties in these systems are listed in Table \ref{table_0}. 

We employed one of the latest physics-based models, compatible with recent satellite solar observations (see \citeads{2007Sci...318.1574D}; \citeads{2011Natur.475..477M}), and currently used for space weather forecast in the solar system (see \citeads{2012JCoPh.231..870T}). This model considers a data-driven approach where surface magnetic field distributions (i.e. \textit{magnetograms}) are decomposed in a high-degree spherical harmonics expansion and implemented as an initial condition, following the methodology presented in \citeads{2011ApJ...732..102T}. Then the simulation evolves self-consistently, calculating coronal heating and stellar wind acceleration based on Alfv\'en wave turbulence dissipation (\citeads{2013ApJ...764...23S}; \citeads{2014ApJ...782...81V}). 

%Table from next section.
\begin{table}
\caption{Basic observational properties of the considered systems.}             
\label{table_0}      
\centering
{\small          
\begin{tabular}{l c c c c c c c}    
\hline\hline
& & & & & & \\[-9pt]                  
Star ID &  $M_{*}$ & $R_*$ & $P_{\rm rot}$ & $M_{\rm p}\sin i$ & $a$ & $e$ \\
 & [$M_{\odot}$] &  [$R_{\odot}$] & [days] & [$M_{\jupiter}$] & [AU] &\\
\hline 
& & & & & & \\[-8pt]                   
HD 1237\,\tablefootmark{a} & 0.86 & 1.00 & 7.00 & 3.37 & 0.49 & 0.51\\  
HD 22049\,\tablefootmark{b,$\dagger$} & 0.74 & 0.86 & 11.68 & 1.05 & 3.38 & 0.25\\  
HD 147513\,\tablefootmark{c} & 0.98 & 1.07 & 10.00 & 1.21 & 1.32 & 0.26\\  
\hline                  
\end{tabular}}
\tablefoot{References for each system: \tablefoottext{a}{\citetads{2001A&A...375..205N}; \citetads{2010ApJ...720.1290G}; \citetads{2015A&A...582A..38A}.} \tablefoottext{b}{\citetads{1993ApJ...412..797D}; \citetads{1996ApJ...466..384D}; \citetads{2005ApJS..159..141V}; \citetads{2006ApJ...646..505B}.} \tablefoottext{c}{\cite{2004A&A...415..391M}, \citetads{2007ApJS..168..297T}; \citeads{2016A&A...585A..77H}.} \\\tablefoottext{$\dagger$}{The listed orbital parameters in \citetads{2016A&A...588A..28A} for the exoplanet in this system were taken from the discovery paper (\citeads{2000ApJ...544L.145H}).}}
\end{table}

For other stars this information can be nowadays retrieved (to some extent) from high-resolution spectropolarimetric observations and the technique of Zeeman-Doppler imaging (ZDI, \citeads{1989A&A...225..456S}; \citeads{1991A&A...250..463B}; \citeads{1997A&A...326.1135D}; \citeads{2000MNRAS.318..961H}; \citeads{2002A&A...381..736P}). For the stellar systems of interest, ZDI maps were previously recovered by \citetads{2014A&A...569A..79J}, \citetads{2015A&A...582A..38A}, and \citetads{2016A&A...585A..77H}. As discussed in detail in \citetads{2016A&A...588A..28A}, we considered two different implementations of this mapping technique (i.e. ZDI and SH-ZDI). The first one considers an image reconstruction with independent pixels for the radial, meridional, and azimuthal components, yielding an unconstrained distribution of the magnetic field (\citeads{1991A&A...250..463B}; \citeads{1997A&A...326.1135D}). The second one follows the methodology presented in \citetads{2001MNRAS.322..681H} and \citetads{2006MNRAS.370..629D}, where the vector field is reconstructed in terms of a spherical harmonics decomposition. The main difference with the previous approach is the possibility to impose physical and geometrical constraints to the final reconstructed topology (i.e. purely potential or toroidal fields, symmetry or antisymmetry). In this way, the SH-ZDI implementation permits the completion of the map in the un-observed hemisphere (due to the inclination of the star). As explained in \citetads{2016A&A...588A..28A}, the SH-ZDI maps were completed preserving the same level of fit to the spectropolarimetric observations as the one obtained with the standard ZDI procedure. This last consideration is fundamental due to the fact that the field strengths in the final map depend on the degree of fit to the observations. This ZDI\,/\,SH-ZDI comparison was performed in order to evaluate the effect of these observational constraints in our simulations. Additionally, the analysis included a comprehensive evaluation procedure of two solar simulations against satellite data, covering activity minimum (CR\,1922) and maximum (CR\,1962), that were spatially-filtered to a similar level of resolution as the ZDI maps. Finally, the results of the ZDI-driven stellar simulations were consistently compared with the solar cases, and with various observational estimates of their coronal conditions. 

This paper provides continuity to this previous work, to include the stellar wind properties and inner astrospheric structure. A description of the numerical set up, boundary conditions, and general characteristics in each simulation domain is provided in Sect. \ref{sec_model}. Section \ref{sec_results} contains the results for each system, including the reference solar cases (i.e., activity minimum and maximum). We discuss our results in the context of previous observational and numerical studies in Sect. \ref{sec_Analysis}, and the main conclusions of our work are summarised in Sect. \ref{sec_Summary}. 

\section{3D MHD simulation: Winds and inner astrospheres}\label{sec_model}

As in the first paper of this study, the numerical simulations have been performed using the three-dimensional MHD code BATS-R-US (\citeads{1999JCoPh.154..284P}) as part of the Space Weather Modeling Framework (SWMF, T\'oth et al. \citeyearads{2005JGRA..11012226T}, \citeyearads{2012JCoPh.231..870T}). We consider the Alfv\'en Wave Solar Model (AWSoM), which solves the two-temperature MHD equations with additional pressure and energy terms associated with the propagation, reflection, and transmission of low-frequency Alfv\'en waves. The complete description of the code and its numerical implementation can be found in \citetads{2014ApJ...782...81V}.  

In this case we analyse the properties of the stellar wind, including the inner region of the stellar corona (SC module) and the resulting structure in the inner astrosphere (IH module). The solar\,/\,stellar cases, and the definition of the entire SC component, are identical as in \citetads{2016A&A...588A..28A}. This includes base conditions typically assumed in high-resolution solar simulations, to match solar observations such as \emph{in-situ} wind properties at $1$~AU and line-of-sight EUV/X-ray images (\citeads{2013ApJ...764...23S}; \citeads{2013ApJ...778..176O}). As ZDI reconstructions necessarily have limited spatial resolution because they are insensitive to the small-scale surface field, we used solar magnetograms that have been spatially-filtered to a comparable resolution as the ZDI/SH-ZDI maps. This allows us to better quantify the effects of this limitation in our simulations consistently (see also \citeads{2016A&A...588A..28A}). As previously discussed by \citetads{2013ApJ...764...32G}, we expect a much smaller effect in the wind structure than in the X-ray morphology.

The IH component covers the domain from $25\,R_*$ up to $215\,R_*$ ($\sim1$\,AU in solar units). The physical conditions in this domain are calculated in the ideal MHD regime, and the simulations have been driven by coupling the steady-state stellar corona solutions described in the previous paper of this study, as the inner boundary conditions of the astrospheric component. A $5\,R_*$ domain-overlap (from $25\,R_*$ to $30\,R_*$) is used in the coupling procedure between both domains\footnote[3]{More details are available in \url{http://csem.engin.umich.edu/tools/swmf/documentation/HTML/SWMF/index.html}.}

For the simulations of HD 1237, the inner astrospheric solution is additionally coupled as a boundary condition of the global magnetosphere (GM) module of the SWMF\footnote[4]{See \url{http://ccmc.gsfc.nasa.gov/cgi-bin/display/RT_t.cgi?page=mpause} for real-time monitoring of the Earth's magnetosphere, using satellite data and the SWMF.}. The spatial locations used for this coupling are described in Sect. \ref{sec_HD1237b}. Inside the GM module, the boundary is set at $50\,R_{\rm p}$ (planet radii) towards the central star (day side). The spatial domain for this module extends up to $150\,R_{\rm p}$ in the night side and to $75\,R_{\rm p}$ in the perpendicular directions (orthogonal to the star-planet axis). Both simulation domains (IH and GM components), use a non-uniform cartesian grid which is automatically refined at the location of large gradients either of the magnetic field or plasma density. In this way the geometry of the astrospheric current sheet (in the IH module), and the bow shock structure (in the GM module) are properly resolved.

The inner boundary condition (located at $\sim$\,1\,$R_{\rm p}$), is defined by the planetary parameters of mass, radius, and dipolar magnetic field strength. The mass is taken from the orbital solution determined by \citetads{2001A&A...375..205N}. There are no observational constraints for the remaining two parameters. A Jupiter-size planet ($R_{\rm p} = R_{\jupiter}$), with a fiducial dipolar field of $B_p = 1$ G, is assumed in this case. Previous observational and numerical studies focused on hot-Jupiter systems, have suggested stronger planetary magnetic fields (e.g. Shkolnik et~al. \citeyearads{2005ApJ...622.1075S}, \citeyearads{2008ApJ...676..628S}; Vidotto et~al. \citeyearads{2010ApJ...722L.168V}, \citeyearads{2012MNRAS.423.3285V}; Llama et~al. \citeyearads{2011MNRAS.416L..41L}, \citeyearads{2013MNRAS.436.2179L}). However, the selection of a stronger (weaker) planetary magnetic field would mainly lead to a larger (smaller) size of the magnetosphere, following the relation $R_{\rm M} \propto B_{\rm p}^{1/3}$ (Eq. \ref{eq_2}, Sect. \ref{sec_HD1237b}). As we are interested in characterising the relative effects from the resulting stellar wind structure (and its connection with the magnetic field topology at the stellar surface), these are independent from the assumed field strength for the exoplanet. 

\section{Numerical results}\label{sec_results}

\noindent The results for each simulation regime are presented in the following sections. From the solution in the SC module we computed the mass and angular momentum loss rates (Sect. \ref{sec_alfven}). Inside the IH module, the steady-state solution led to the global structure of the wind and associated current sheet (Sect. \ref{sec_resultsIH}). As in \citetads{2016A&A...588A..28A}, we perform a consistent comparison between the solar and stellar cases. This includes the solar minimum and maximum activity states, as well as the ZDI/SH-ZDI driven simulations for the stellar cases. Finally, the results of HD 1237 are presented in Sect. \ref{sec_HD1237b}, including the coupled solutions of the GM module, at two critical locations of the exoplanetary orbit.

\subsection{Alfv\'en surface, mass and angular momentum loss rates}\label{sec_alfven}

\begin{figure*}[!ht]
\centering %  left, bottom, right and top
\includegraphics[trim=0.3cm 2.5cm 0.3cm 0.0cm, clip=true, width=\hsize/3]{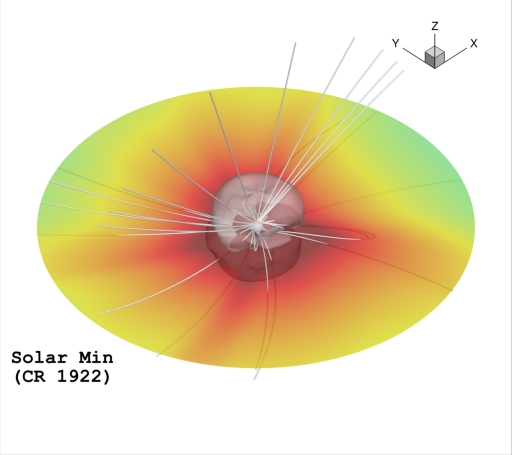}\includegraphics[trim=0.3cm 2.5cm 0.3cm 0.0cm, clip=true, width=\hsize/3]{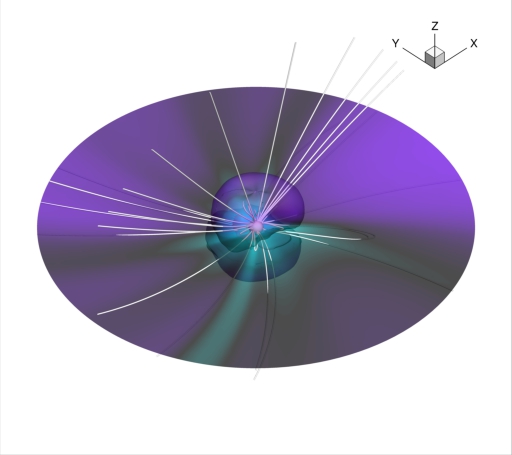}\includegraphics[trim=0.3cm 2.5cm 0.3cm 0.0cm, clip=true, width=\hsize/3]{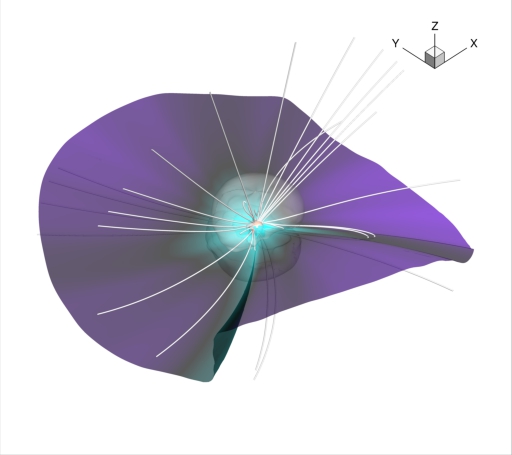}

\includegraphics[trim=0.3cm 0.0cm 0.3cm 0.0cm, clip=true, width=\hsize/3]{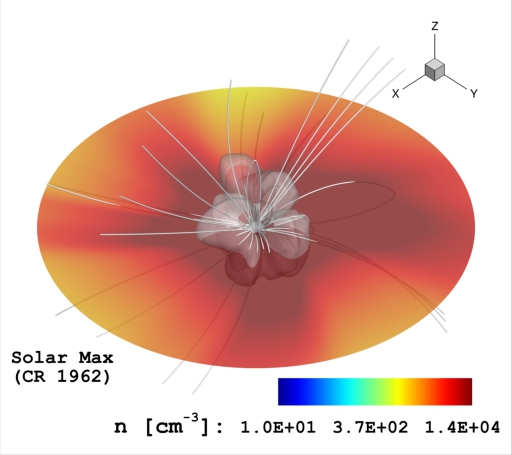}\includegraphics[trim=0.3cm 0.0cm 0.3cm 0.0cm, clip=true, width=\hsize/3]{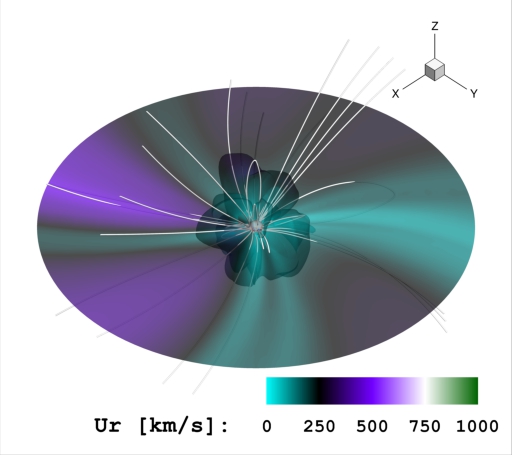}\includegraphics[trim=0.3cm 0.0cm 0.3cm 0.0cm, clip=true, width=\hsize/3]{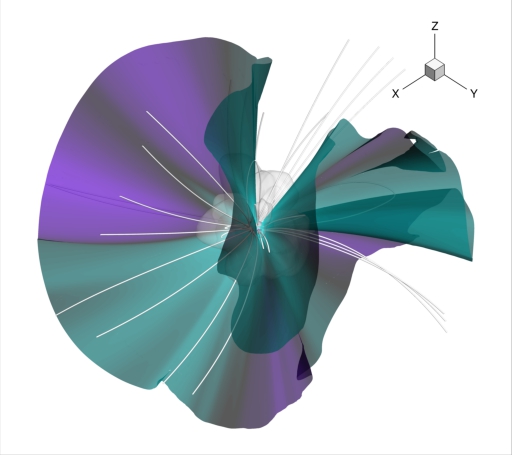}
\caption{Simulation results in the solar corona (SC) domain for activity minimum (CR 1922, up) and maximum (CR 1962, down). The left and middle panels contain the projection onto the equatorial plane ($z = 0$) of the plasma density $n$, and the radial wind speed $u_{\rm r}$, respectively. In the right panel the distribution of $u_{\rm r}$ on the developed current sheet structure ($B_{\rm r} = 0$) is presented. The translucent shade denotes the Alfv\'en surface ($M_{\rm A} = 1$) calculated from the steady-state solution. The corresponding colour scales for $n$ and $u_{\rm r}$ are preserved among the different panels. Selected 3D magnetic field lines are shown in white.}
\label{fig_1}
\end{figure*}

\begin{figure*}[!ht]
\centering %  left, bottom, right and top
\includegraphics[trim=0.3cm 0.0cm 0.3cm 0.0cm, clip=true, width=\hsize/3]{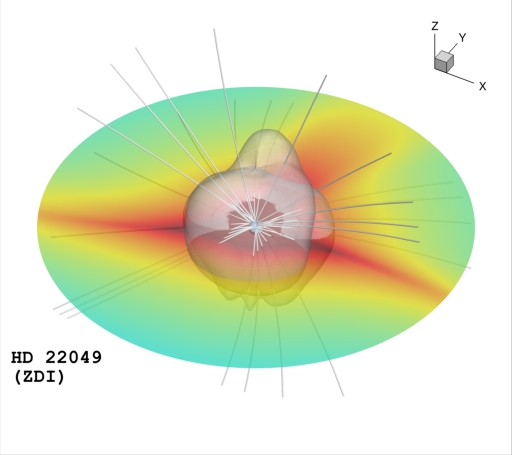}\includegraphics[trim=0.3cm 0.0cm 0.3cm 0.0cm, clip=true, width=\hsize/3]{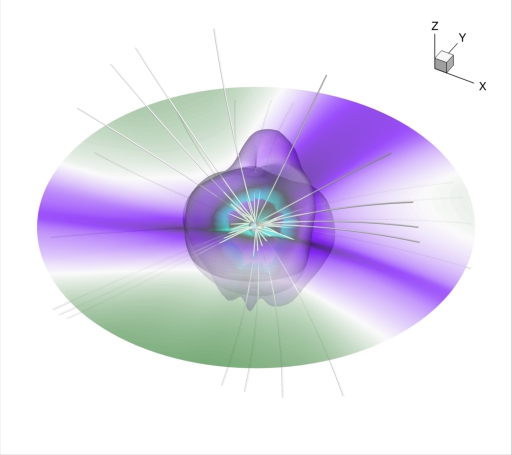}\includegraphics[trim=0.3cm 0.0cm 0.3cm 0.0cm, clip=true, width=\hsize/3]{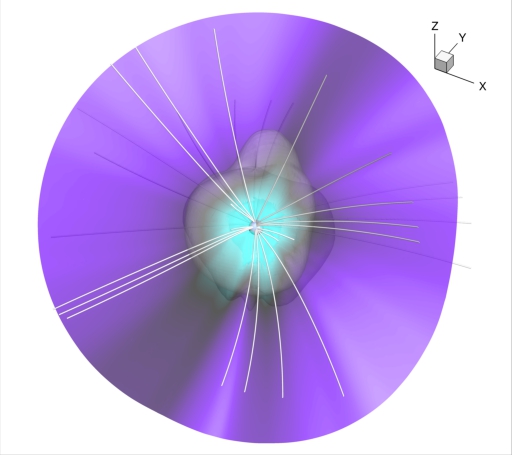}
\includegraphics[trim=0.3cm 0.0cm 0.3cm 0.5cm, clip=true, width=\hsize/3]{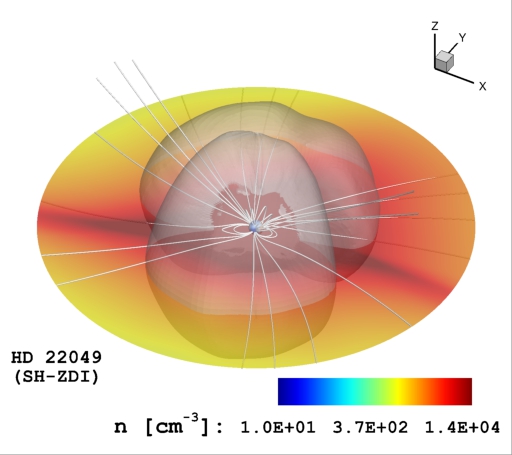}\includegraphics[trim=0.3cm 0.0cm 0.3cm 0.5cm, clip=true, width=\hsize/3]{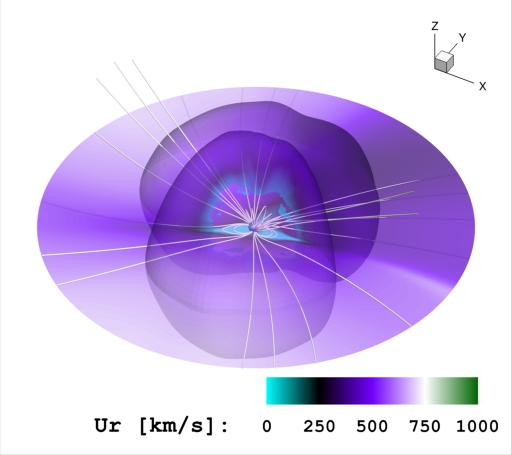}\includegraphics[trim=0.3cm 0.0cm 0.3cm 0.0cm, clip=true, width=\hsize/3]{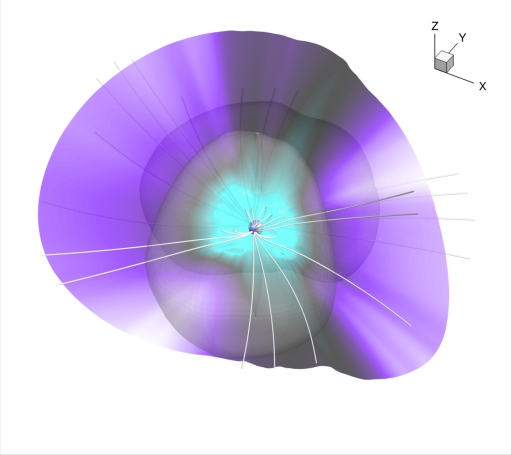}
\caption{Simulation results in the SC domain for HD 22049 driven by the ZDI (up) and SH-ZDI (down) magnetic field maps. The left and middle panels contain the projection onto the equatorial plane ($z = 0$) of the plasma density $n$, and the radial wind speed $u_{\rm r}$, respectively. In the right panel the distribution of $u_{\rm r}$ on the developed current sheet structure ($B_{\rm r} = 0$) is presented. The translucent shade denotes the Alfv\'en surface ($M_{\rm A} = 1$) calculated from the steady-state solution. The corresponding colour scales for $n$ and $u_{\rm r}$ are preserved among the different panels. Selected 3D magnetic field lines are shown in white.}
\label{fig_2}
\end{figure*}

\begin{figure*}[!ht]
\centering %  left, bottom, right and top
\includegraphics[trim=0.3cm 2.0cm 0.3cm 0.0cm, clip=true, width=\hsize/3]{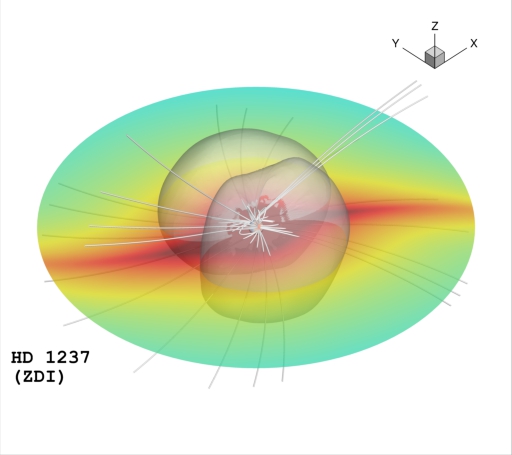}\includegraphics[trim=0.3cm 2.0cm 0.3cm 0.0cm, clip=true, width=\hsize/3]{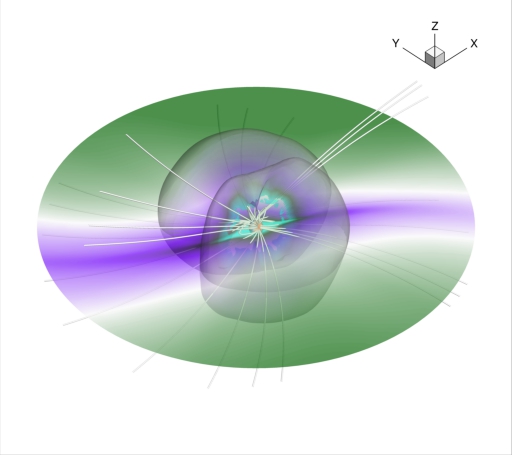}\includegraphics[trim=0.3cm 2.0cm 0.3cm 0.0cm, clip=true, width=\hsize/3]{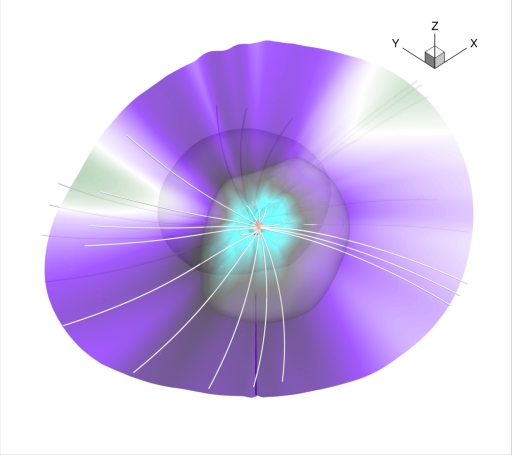}

\includegraphics[trim=0.3cm 0.0cm 0.3cm 0.0cm, clip=true, width=\hsize/3]{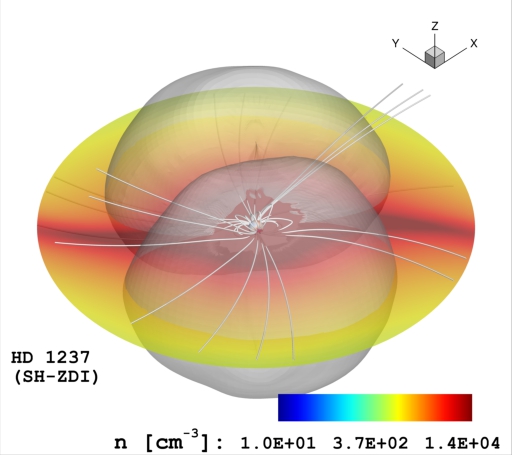}\includegraphics[trim=0.3cm 0.0cm 0.3cm 0.0cm, clip=true, width=\hsize/3]{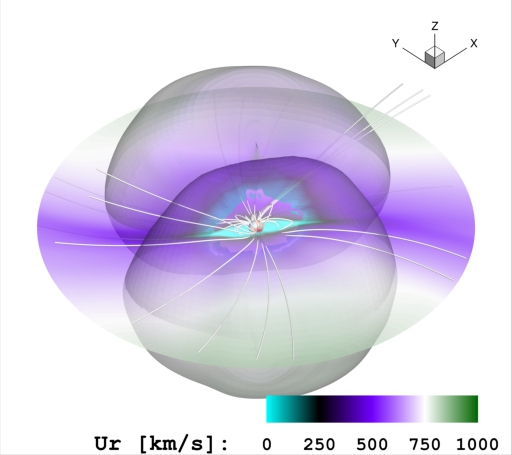}\includegraphics[trim=0.3cm 0.0cm 0.3cm 0.0cm, clip=true, width=\hsize/3]{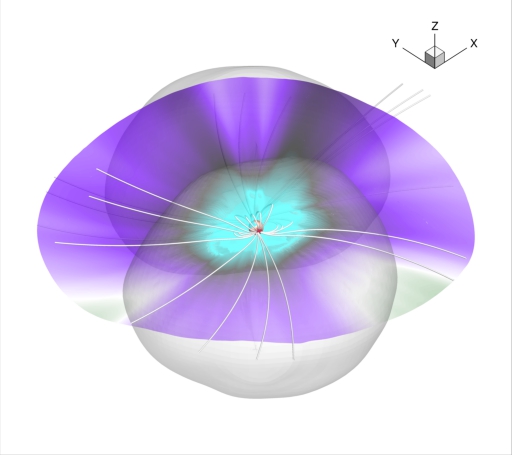}
\caption{Simulation results in the SC domain for HD 1237 driven by the ZDI (up) and SH-ZDI (down) magnetic field maps. The left and middle panels contain the projection onto the equatorial plane ($z = 0$) of the plasma density $n$, and the radial wind speed $u_{\rm r}$, respectively. In the right panel the distribution of $u_{\rm r}$ on the developed current sheet structure ($B_{\rm r} = 0$) is presented. The translucent shade denotes the Alfv\'en surface ($M_{\rm A} = 1$) calculated from the steady-state solution. The corresponding colour scales for $n$ and $u_{\rm r}$ are preserved among the different panels. Selected 3D magnetic field lines are shown in white.}
\label{fig_4}
\end{figure*}

\begin{figure*}[!ht]
\centering %  left, bottom, right and top
\includegraphics[trim=0.3cm 0.0cm 0.3cm 0.0cm, clip=true, width=\hsize/3]{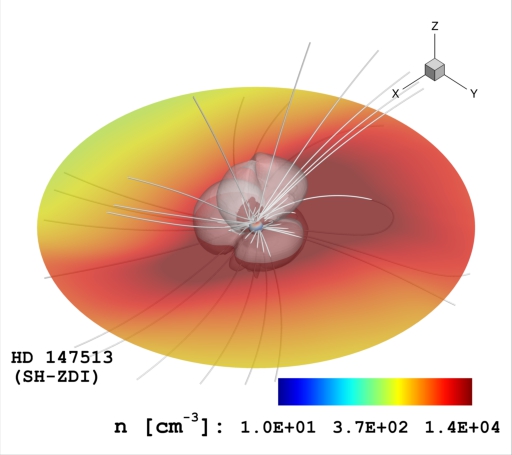}\includegraphics[trim=0.3cm 0.0cm 0.3cm 0.0cm, clip=true, width=\hsize/3]{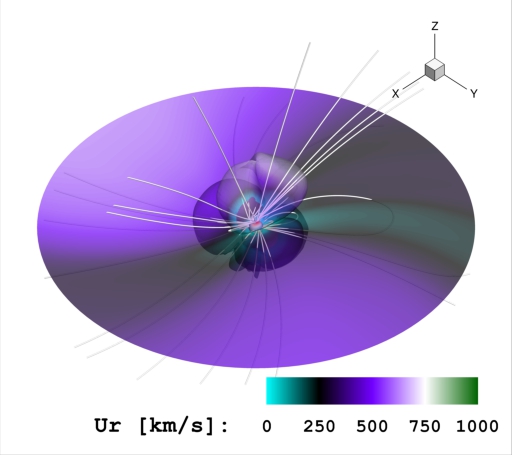}\includegraphics[trim=0.3cm 0.0cm 0.3cm 0.0cm, clip=true, width=\hsize/3]{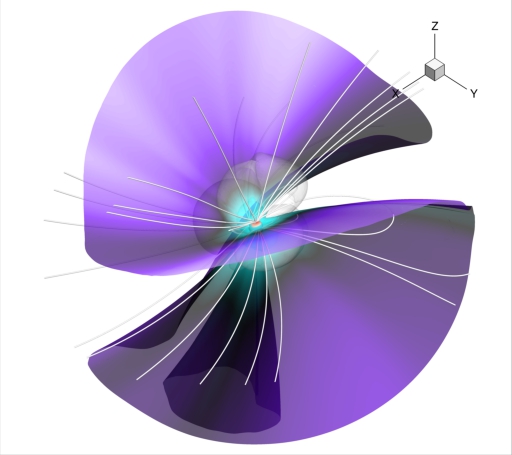}
\caption{Simulation results in the SC domain for HD 147513 driven by the SH-ZDI magnetic field map. See caption of Fig. \ref{fig_4}.}
\label{fig_3}
\end{figure*}

\noindent We initially describe the properties of the solar/stellar wind inside the SC module.  Figures \ref{fig_1} to \ref{fig_4} show the equatorial distribution of the plasma density $n$, and radial wind speed $u_{\rm r}$ (left and middle panels, respectively), extracted from the corresponding steady-state solutions. The right panels contain the distribution of $u_{\rm r}$ over the current sheet structure, which is defined as the iso-surface with $B_{\rm r} = 0$. We also compute the resulting Alfv\'en surface (AS) for each solution, displayed as a translucent shade in Figs. \ref{fig_1} to \ref{fig_4}. This is performed by calculating the spatial locations at which the Alfv\'enic Mach number $M_{\rm A} = u_{\rm sw} / v_{\rm A} = 1$. In this relation $u_{\rm sw}$ represents the local stellar wind speed, while $v_{\rm A}$ is the Alfv\'en speed of the plasma, determined by the ratio $B\,/\sqrt{4\pi\rho}$, with $B$ and $\rho$ as the local magnetic field strength and density, respectively. 

Figure \ref{fig_1} shows the results of the solar simulations in this domain. For both activity states, the expected global thermodynamical properties of the solar wind are achieved; relatively fast and low density during solar minimum, and considerably slower and denser for activity maximum. Additionally, as evidenced by the developed current sheet (right panel in Fig. \ref{fig_1}) and the geometry of the AS, the overall complexity of the driving  magnetic field distribution is reflected in the wind solution. During activity minimum, the current sheet is mostly confined to the equatorial plane (with small deviations in particular sectors of the structure). The AS shows a two-lobe structure aligned with the rotation axis of the star (z-axis), which is usually obtained for simple (nearly dipolar) surface magnetic field distributions (e.g. \citeads{2014MNRAS.438.1162V}; \citeads{2014ApJ...783...55C}). For activity maximum, the structure of the current sheet shows warped sectors and greatly departs from the equatorial plane. Similarly the resulting AS in this case shows multiple lobes of irregular sizes, without any preferred orientation in the 3D domain. These fundamental differences are clearly seen in the resulting structure of the inner heliosphere, presented in Sect. \ref{sec_resultsIH}.  

The simulations of HD 22049 and HD 1237 led to similar wind structures in the SC domain (Figs. \ref{fig_2} and \ref{fig_4}, respectively). Streamers can be observed on the equatorial plane (three for HD 22049 and two for HD 1237), with a mean density $n$\,$\simeq$\,10$^{3}$ cm$^{-3}$ and radial speeds of $u_{\rm r}$\,$\sim$\,500 km s$^{-1}$ in the ZDI-driven cases. For the SH-ZDI simulation of HD 22049 two of these streamers are merged, creating a broader high-density sector on the equatorial plane. For both stars, the density drops by a factor of $\gtrsim$\,100 between the streamers, while the velocity rises to $u_{\rm r}$\,$\sim$\,1000 km s$^{-1}$ and 750 km s$^{-1}$ for the ZDI and the SH-ZDI simulations, respectively. Similarly, a two-lobe AS structure is developed in all the simulations, which is consistently larger in the SH-ZDI cases (see Table \ref{table_1}). The alignment of the AS lobes deviates significantly from the rotation axis, leading to a current sheet structure nearly confined to a plane, highly inclined with respect to the projected stellar equator (Figs. \ref{fig_2} and \ref{fig_4}, right panels). The velocity of the wind along the current sheet is approximately 500 km s$^{-1}$ in all cases, with variations up to $\pm40\%$ in very small locations of the structure. These structures show a higher density ($n$\,$\simeq$\,10$^{4}$ cm$^{-3}$) in the SH-ZDI simulations, with roughly the same velocity as in the ZDI cases.

Compared to the previously described stellar cases, HD 147513 showed a rather different wind structure in this domain (Fig. \ref{fig_3}). The high-density structures of the wind are much wider in this case, with associated radial speeds of $u_{\rm r}$\,$\sim$\,250 km s$^{-1}$. The velocity of the wind remains below 750 km s$^{-1}$ in the equatorial plane, and barely reaches this value in few locations of the 3D domain. Similar to the solar maximum case, the current sheet structure of HD 147513 shows warped sectors and clearly deviates from a planar structure (right panel of Fig. \ref{fig_3}). This additional complexity can be also seen in the  irregular lobes developed in the AS, which are also common with the solar simulation during activity maximum (Fig. \ref{fig_1}, bottom).
   
These results clearly show the importance of the AS properties on the resulting wind structure. By definition, the AS corresponds to the boundary between magnetically-coupled outflows ($M_{\rm A} < 1$) and the escaping stellar wind which no longer exerts torque on the star ($M_{\rm A} > 1$). For this reason it is commonly used in modelling studies to calculate the mass loss rate, $\dot{M}$, and the angular momentum loss rate, $\dot{J}$, associated with the stellar wind (e.g. \citeads{2010ApJ...721...80C}; \citeads{2014ApJ...783...55C}; \citeads{2015ApJ...813...40G}). Furthermore, as illustrated in Figs. \ref{fig_1} to \ref{fig_4}, the topology of the AS reflects to some extent the complexity of the magnetic field distribution driving the simulation (see also Vidotto et al. \citeyearads{2014MNRAS.438.1162V}, \citeyearads{2015MNRAS.449.4117V}; \citeads{2015ApJ...807L...6G}), which in turn, in a self-consistent model, should be directly related to the resulting coronal structure (see \citeads{2016A&A...588A..28A}). Finally, previous studies of planet-hosting stars with close-in exoplanets have also shown the importance of the exoplanet location with respect to the AS, which can lead to strong magnetic interactions and angular momentum transfer between the star and the planet (\citeads{2014ApJ...790...57C}; Strugarek et al. \citeyearads{2014ApJ...795...86S}, \citeyearads{2015ApJ...815..111S}). 

In addition, the AS provides a common framework to consistently compare our simulations in this domain, and to place our results in context with other studies in the literature. Table \ref{table_1} contains a summary of the resulting stellar wind properties averaged over the AS, as well as the $\dot{M}$ and $\dot{J}$ values in each case. Several important results are obtained from this quantitative analysis. First of all, by taking an average of the activity minimum and maximum cases, we obtain a mean solar mass loss rate <$\dot{M}_{\odot}$>\,=\,$3.78\times10^{-14}$ $M_{\odot}$ yr$^{-1}$, consistent with the nominal accepted value of $\dot{M}_{\odot}$\,$\simeq$\,$2\times\,10^{-14}$ $M_{\odot}$ yr$^{-1}$ (\citeads{2004LRSP....1....2W} and references therein), and the observed scatter during the course of the activity cycle (by a factor of $\sim\,$2, \citeads{2011MNRAS.417.2592C}). 

If we analyse the solar simulations independently, the predicted mass loss rate during activity minimum agrees well with Voyager II data, averaged over the corresponding period of time at the spacecraft (i.e. $\sim$ 9 months after\footnote[2]{Approximate time required for the solar wind to reach Voyager II location, at the average speed predicted by the model in this epoch.} the CR 1922 of May 1997). However, a similar comparison for the considered activity maximum epoch (i.e. $\sim$ 14 months after the CR 1962 of Apr-May 2000), indicates that the solar mass loss rate in this case is overestimated by $\sim$40\% (\citeads{2011MNRAS.417.2592C}). This additional mass escaping the star can be interpreted as a deficit of confining loops in the lower corona (inside the AS), which results from the (spatial) resolution-limited magnetograms driving the simulation (see \citeads{2016A&A...588A..28A}). While this condition was common among both solar simulations, the effect on the activity maximum case is larger, given the relative amount of complexity and magnetic flux lost in the process of spatially degrading the surface field distribution. 

In a similar manner, the simulations yield an average solar angular momentum loss <$\dot{J}_{\odot}$>\,=\,$7.5\times10^{29}$ erg. Unlike the observational estimates of $\dot{M}_{\odot}$, values of $\dot{J}_{\odot}$ are more uncertain (ranging between $\sim$10$^{29}$ -- 10$^{31}$ erg), and usually determined via numerical models with different assumptions (e.g. \citeads{2010ApJ...721...80C}; \citeads{2012ApJ...754L..26M}; \citeads{2014ApJ...783...55C}; \citeads{2015ApJ...813...40G}). The relatively small <$\dot{J}_{\odot}$> resulting from our simulations reflects the average size of the AS directly , which is known to increase with the field strength and decrease with the field complexity (see \citeads{2015ApJ...798..116R}; \citeads{2015ApJ...807L...6G}). In the considered solar cases these dependencies are partially compensated, leading to a similar value of $\left<R_{\rm AS}\right>$ in both simulations (see Table \ref{table_1}).
 
Our result for $\left<R_{\rm AS}\right>$ in the solar minimum case, is very similar to the value obtained by the 2.5-dimensional simulations of \citetads{2015ApJ...798..116R} using the MHD PLUTO code (\citeads{2007ApJS..170..228M}) and a similar activity epoch. Interestingly, for activity maximum we obtain a $\left<R_{\rm AS}\right>$ value which is roughly twice compared to their findings. This difference could be related (among with other possibilities), with the dimensional reduction of their approach, with specific properties of the driving magnetic field distribution (i.e. Carrington rotation and instrument used to map the magnetic field), and with the amount of small scale field included in each numerical implementation. 

Unlike the average size of the AS, the fundamental properties of solar wind at this region show large variations between both activity states; differences by a factor of $\sim$\,6 in the mean plasma density $\left<n\right>_{\rm AS}$, and by a factor of $\sim$\,2 for the average radial wind speed $\left<u_{\rm r}\right>_{\rm AS}$. Smaller differences are obtained for the remaining solar wind parameters. 

For the considered stellar systems, important differences arise between the ZDI and SH-ZDI cases. For the mass and angular momentum loss rates, the resulting values of $\dot{M}$ and $\dot{J}$ of HD 1237 differ by a factors of $\sim$\,3 and $\sim$\,9 respectively, being larger in the SH-ZDI case. Similarly, the SH-ZDI simulation of HD 22049 yields $\dot{M}$ and $\dot{J}$ values which are several times larger than in the corresponding ZDI case (i.e. by factors of $\sim$\,4 and $\sim$\,11, respectively). Smaller differences arise in the average size of the AS, $\left<R_{\rm AS}\right>$, being roughly $1.5$ times larger in the SH-ZDI simulations of these two systems. The obtained differences in $\dot{J}$ appear mainly as a result of its direct dependancy with $\dot{M}$ (see \citeads{2015ApJ...813...40G}). On the other hand, the mass loss rate variations are connected to the relative differences in the surface field distributions driving the simulations (see \citeads{2016A&A...588A..28A}), and the Alfv\'en wave energy transfer to the corona and wind implemented in the model. The latter is described via the Poynting flux of the emerging Alfv\'en waves, $S_{\rm A}$, taken to be proportional to the field strength (and polarity) at the inner boundary of the simulation (i.e., $S_{\rm A} \propto B_{\rm r}$; see \citeads{2014ApJ...782...81V}). Previous studies based on the same MHD solver but with a different wind model (i.e., a thermally-driven polytropic stellar wind), do not display significant variations in $\dot{M}$ when considering changes in the magnetic field geometry and/or the incompleteness of the ZDI maps (see \citeads{2012MNRAS.423.3285V}, which is the basis for the models presented in Vidotto et al. \citeyearads{2014MNRAS.438.1162V}, \citeyearads{2015MNRAS.449.4117V}, \citeads{2016ApJ...820L..15D} and \citeads{2016MNRAS.459.1907N}). As the wind-driving mechanism in our simulations is based on Alfv\'en-wave turbulence dissipation, we find much stronger differences in these wind properties based on the large-scale magnetic field geometry. This is indicative of a radical difference between these other models and our simulations.  

\begin{table*}[!ht]
\caption{Mass loss rates, $\dot{M}$ and angular momentum loss rates, $\dot{J}$, calculated from the steady-state solutions. The additional stellar wind properties represent averages over the resulting Alfv\'en surface (AS), while $\left<R_{\rm AS}\right>$ corresponds to the mean AS radius in each case.}             
\label{table_1}      
\centering
{\normalsize          
\begin{tabular}{ c | c c | c c | c | c c }    
\hline\hline
Parameter &  \multicolumn{2}{|c|}{HD 1237} & \multicolumn{2}{|c|}{HD 22049} & HD 147513 &  \multicolumn{2}{|c}{Sun}\\
 & ZDI & SH-ZDI & ZDI & SH-ZDI & SH-ZDI & CR 1922 (Min) & CR 1962 (Max) \\
\hline
& & & & & & &\\[-9pt] 
$\dot{M}$ [$\times\,10^{-14}$ $M_{\odot}$ yr$^{-1}$] & 4.70 & 13.9 & 2.77 & 10.2 & 11.4 & 2.76 & 4.80 \\
$\dot{J}$ [$\times\,10^{30}$ erg] & 6.77 & 58.0 & 1.10 & 12.3 & 3.66 & 0.40 & 1.10 \\
$\left<u_{\rm r}\right>_{\rm AS}$ [km s$^{-1}$] & 679 & 654 & 562 & 493 & 363 & 364 & 171 \\
$\left<n\right>_{\rm AS}$ [$\times\,10^{4}$ cm$^{-3}$] & 2.85 & 3.09 & 2.65 & 3.64 & 15.1 & 4.97 & 30.0 \\
$\left<T\right>_{\rm AS}$ [$\times\,10^{6}$ K] & 2.33 & 1.52 & 1.81 & 1.34 & 1.54 & 1.26 & 0.84 \\
$\left<B\right>_{\rm AS}$ [$\times\,10^{-2}$ G] & 1.94 & 2.08 & 1.82 & 2.23 & 2.98 & 1.64 & 2.63 \\
$\left<R_{\rm AS}\right>$ [$R_{*}$] & 12.0 & 19.6 & 10.2 & 15.6 & 7.1 & 6.4 & 6.8 \\[1pt]
%$N$ & 2 & 2 & 2 & 2 &  &  & \\
\hline                  
\end{tabular}}
\end{table*}

\begin{figure*}[!ht]
\centering %  left, bottom, right and top
\includegraphics[trim=0.2cm 0.0cm 0.2cm 0.0cm, clip=true,scale=0.42]{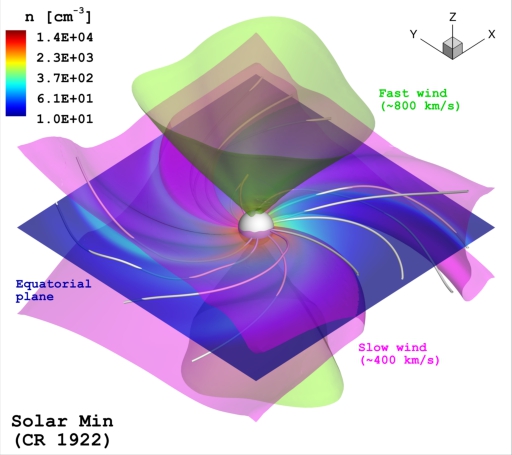}\hspace{0.8cm} \includegraphics[trim=0.2cm 0.0cm 0.2cm 0.0cm, clip=true, scale=0.42]{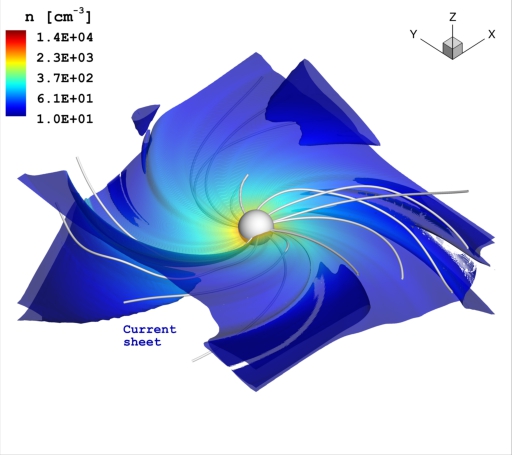}

\includegraphics[trim=0.2cm 0.0cm 0.2cm 0.0cm, clip=true, scale=0.42]{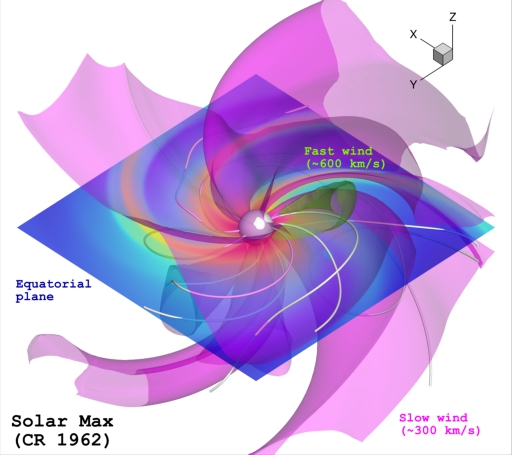}\hspace{0.8cm} \includegraphics[trim=0.2cm 0.0cm 0.2cm 0.0cm, clip=true, scale=0.42]{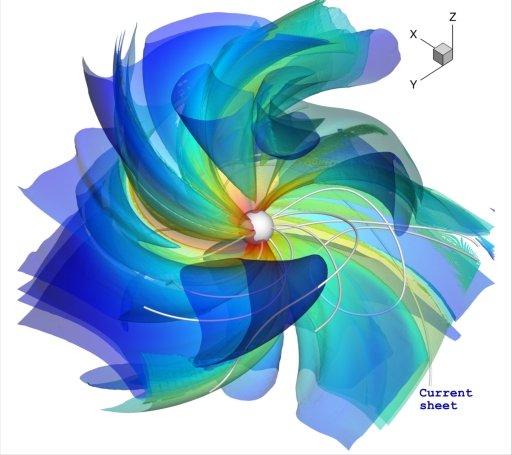}
\caption{Simulation results in the inner heliosphere (IH) domain for activity minimum (CR 1922, top) and maximum (CR 1962, bottom). The central white sphere denotes the boundary with the solar corona (SC) domain at 25\,$R_{\odot}$ (Sect. \ref{sec_alfven}). The density structure of the steady-state solution is displayed on the equatorial plane (left) and the heliospheric current sheet (right). In the left panel, the topology and associated magnitudes of the dominant radial velocity components ($u_{\rm r}$) of the solar wind are also included (fast: green -- slow: magenta). The density ($n$) colour scale is preserved among the different panels. Selected 3D magnetic field lines are shown in white.}
\label{fig_5}
\end{figure*}

The remaining stellar wind properties (averaged over the AS), showed less variation between the ZDI and the SH-ZDI cases (see Table \ref{table_1}). Assuming the same initial base conditions, the SH-ZDI simulations led to denser (by\,$\sim$10\,--\,30\%) and colder (by\,$\sim$25\,--\,35\%) winds compared to the ZDI-driven cases. As with the solar simulations, the average stellar wind speed at the AS seems to be lower for higher surface field strengths. These differences are related to the radial behaviour of the thermodynamical quantities (i.e. $\left<R_{\rm AS}\right>$ is larger in the SH-ZDI case), and the underlying coronal structure, which in turn depends on additional factors such as the ZDI map resolution, completeness, and field complexity (see \citeads{2016A&A...588A..28A}). This clearly shows the importance of numerical models which self-consistently simulate both, the corona and stellar wind domains. In the following section, we present the resulting solar and stellar wind properties inside the IH module.

\begin{figure*}[!t]
\centering %  left, bottom, right and top
\includegraphics[trim=0.2cm 0.0cm 0.2cm 0.0cm, clip=true, scale=0.51]{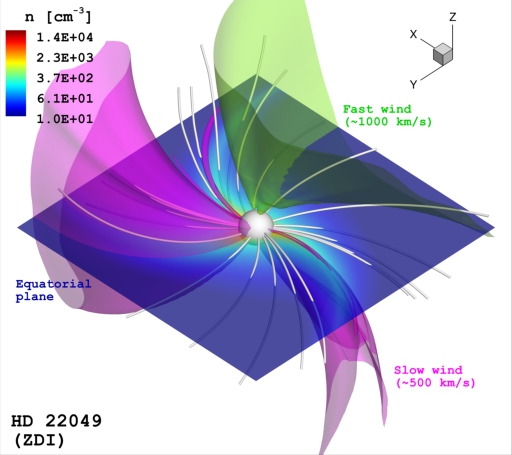}\hspace{0.2cm} \includegraphics[trim=0.2cm 0.0cm 0.2cm 0.0cm, clip=true, scale=0.51]{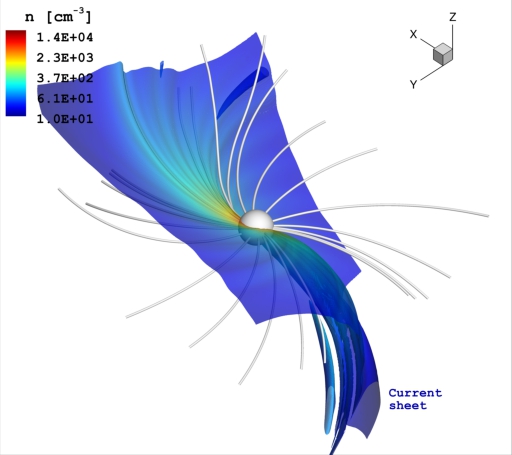}

\includegraphics[trim=0.2cm 0.0cm 0.2cm 0.0cm, clip=true, scale=0.51]{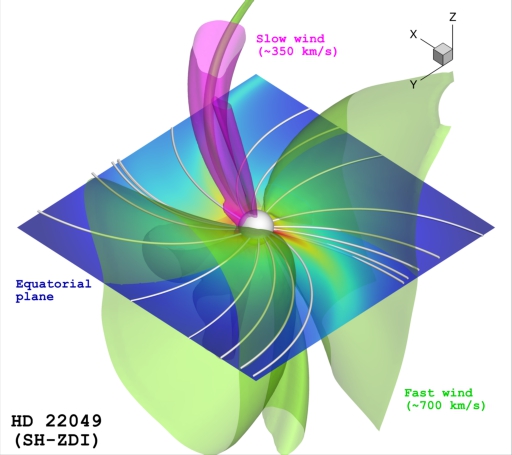}\hspace{0.2cm} \includegraphics[trim=0.2cm 0.0cm 0.2cm 0.0cm, clip=true, scale=0.51]{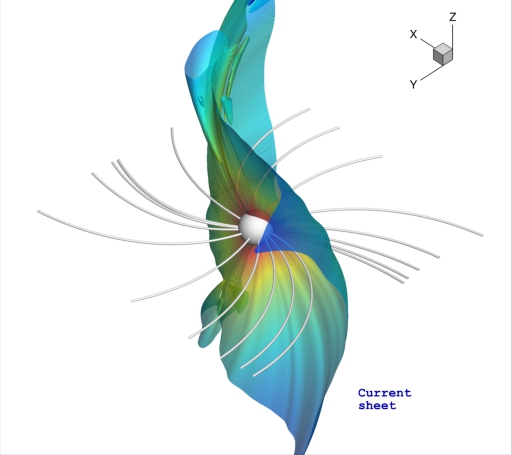}
\caption{Simulation results in the IH domain for HD 22049 driven by the ZDI (top) and SH-ZDI (bottom) magnetic field maps. The central white sphere denotes the boundary with the SC domain at 25\,$R_{*}$ (Sect. \ref{sec_alfven}). The density structure of the steady-state solution is displayed on the equatorial plane (left) and the astrospheric current sheet (right). In the left panel, the topology and associated magnitudes of the dominant radial velocity components ($u_{\rm r}$) of the stellar wind are also included (fast: green -- slow: magenta). The perspective and density colour scale are preserved among the different panels. Selected 3D magnetic field lines are shown in white.}
\label{fig_6}
\end{figure*}

\subsection{Stellar winds and astrospheric current sheet}\label{sec_resultsIH}

\begin{figure*}[!ht]
\centering %  left, bottom, right and top
\includegraphics[trim=0.2cm 0.0cm 0.2cm 0.0cm, clip=true, scale=0.51]{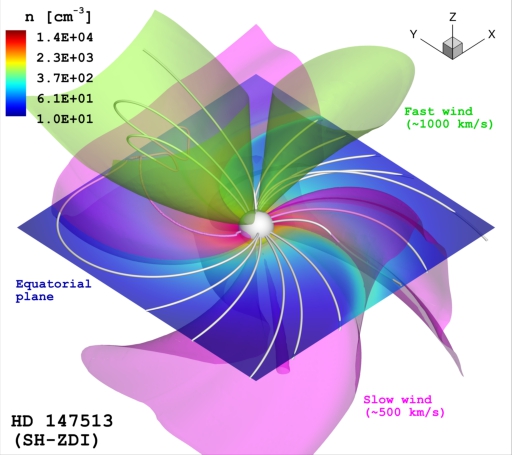}\hspace{0.2cm} \includegraphics[trim=0.2cm 0.0cm 0.2cm 0.0cm, clip=true, scale=0.51]{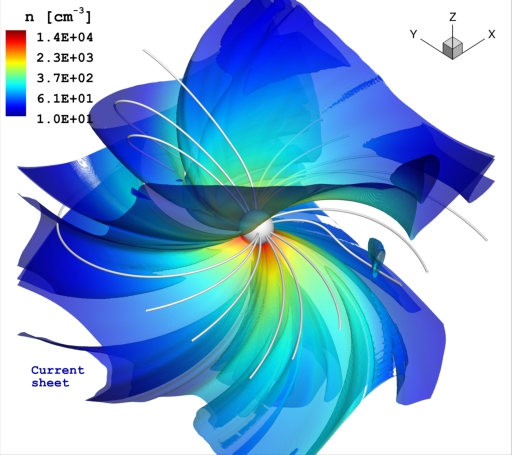}
\caption{Simulation results in the IH domain for HD 147513 driven by the SH-ZDI magnetic field maps. The central white sphere denotes the boundary with the SC domain at 25\,$R_{*}$ (Sect. \ref{sec_alfven}). The density structure of the steady-state solution is displayed on the equatorial plane (left) and the astrospheric current sheet (right). In the left panel, the topology and associated magnitudes of the dominant radial velocity components ($u_{\rm r}$) of the stellar wind are also included (fast: green -- slow: magenta). Selected 3D magnetic field lines are shown in white.}
\label{fig_7}
\end{figure*}

\noindent As mentioned in Sect. \ref{sec_model}, the simulations on the IH domain are driven by the steady-state solutions of the SC region, coupled at 25 $R_{*}$ (white sphere in Figs. \ref{fig_5} to \ref{fig_10} and Fig. \ref{fig_12}). Similarly to the SC domain, we present the density structure of the wind, projected onto the equatorial plane, and the heliospheric/astrospheric current sheet. The associated colour scale for $n$ is preserved between the SC and IH results, showing the consistency of the coupled MHD solution. Additionally, the 3D structure of the radial wind velocity is visualised via two iso-velocity surfaces (translucent shades in Figs. \ref{fig_5} to \ref{fig_9}, left), labeled as fast (green) and slow (magenta) wind components. The magnitude of the fast component is calculated using the peak wind velocity achieved in the simulation, and taking the \textit{floor} with respect to a 100 km s$^{-1}$ velocity bin width. The magnitude of the slow wind component is simply taken as half of the previously defined fast wind. As an example, the solar minimum simulation showed a peak wind velocity of $u_{\rm r}$\,$\sim$\,885 km s$^{-1}$, so the fast and slow wind iso-surfaces were taken at 800 km s$^{-1}$ and 400 km s$^{-1}$, respectively (Fig. \ref{fig_5}, top). The 600 km s$^{-1}$ and 300 km s$^{-1}$ solar wind components in the activity maximum case (Fig. \ref{fig_5}, bottom), were defined in the same way, as a result of a peak wind speed of $u_{\rm r}$\,$\sim$\,614 km s$^{-1}$ in the 3D domain.  

As can be seen from Fig. \ref{fig_5}, the expected global properties and topology of the solar wind (for both activity states), are properly recovered in the simulation. The activity minimum case shows the classical solar wind configuration, with the fast wind emerging from the poles, and the slow wind close to the equator, describing a ``ballerina skirt'' shape along the heliospheric current sheet (Fig. \ref{fig_5}, top right). In the activity maximum case, a more complex solution is obtained, with a 3D structure dominated by the slow component which is no longer restricted to lower latitudes. As the overall velocity of the wind is reduced (compared to the solar minimum case), the fast wind component is nearly inexistent in this solution (Fig. \ref{fig_5}, bottom left). Additionally, a dramatic change in complexity can be observed in the heliospheric current sheet (Fig. \ref{fig_5}, bottom right), as was expected from the resulting topology of this structure inside the inner domain of the simulation (Sect. \ref{sec_alfven}). The density structure of the solar wind is clearly enhanced during the activity maximum solution, as can be compared from the equatorial and current sheet projections in Fig. \ref{fig_5}. This is quantified in more detail in Sect. \ref{sec_Analysis}.  

Figure \ref{fig_6} contains the results for HD 22049, driven by the ZDI (top) and the SH-ZDI (bottom) magnetic field maps. As expected from the SC region (Sect. \ref{sec_alfven}), differences in the geometry and the wind properties are developed in this domain\footnote[2]{See also the 3D animations provided as supplementary material.}. Only one broad fast wind region emerges in the ZDI-driven case, with an associated speed of $u_{\rm r}$\,$\sim$\,1000 km s$^{-1}$ (Fig. \ref{fig_6}, top-left). This wind component is roughly perpendicular to the astrospheric current sheet which, close to the star, displays a tilt of $\sim$\,45$^{\circ}$\footnote[3]{With respect to the stellar rotation axis (i.e., $z$-axis).}, and a rotational drag by the wind at farther distances (Fig. \ref{fig_6}, top-right). In turn, four such fast wind regions are formed in the SH-ZDI simulation, displaying a $\sim$\,30\% reduction of the wind speed (Fig. \ref{fig_6}, bottom-left). From this wind regime, the two broader structures are again nearly perpendicular to each side of the astrospheric current sheet, which in this case is almost orthogonal to the equatorial plane (Fig. \ref{fig_6}, bottom-right). The remaining two fast wind regions appear as collimated jet-like structures, closely aligned with the local orientation of astrospheric current sheet. The latter, as with the solar minimum case (Fig. \ref{fig_1}, top), shows a connection with the slow wind region in both simulations of HD 22029, where the denser material is carried away from the star. 

\noindent The results for HD 147513 inside the IH simulation domain are presented in Fig. \ref{fig_7}, where the maximum radial wind speed was $u_{\rm r}$\,$\sim$\,1040 km s$^{-1}$. Two cone-shaped regions, associated with the fast wind component, appear close the north pole of the star. No southern counterpart for these regions was obtained in the simulation. The topologies of the slow wind component and the astrospheric current sheet, clearly resemble the solar maximum solution in this domain (see Fig. \ref{fig_5}, bottom). This is consistent with the results obtained inside the SC module (Sect. \ref{sec_alfven}) and with the simulated global properties of the corona in both cases (\citeads{2016A&A...588A..28A}). However, as discussed in the first paper of this study, these results could be affected by the relatively low spatial resolution of the SH-ZDI map of HD 147513 (see \citeads{2016A&A...585A..77H}). Still, this solution indicates that the coronal structure and wind characteristics may be extremely complex, even for cases with a relatively simple surface field distribution. In this context, scaling relations involving average stellar/magnetic properties and extrapolations, cannot provide complete descriptions of the coronal and wind conditions of a particular system. This is critical for characterising planet-hosting stars, where those specific environmental properties (e.g. coronal emission, stellar wind structure, mass loss, etc) will strongly affect the exoplanetary conditions of the system. 

\subsection{Environment of the HD 1237 system}\label{sec_HD1237b}

\begin{figure*}[!ht]
\centering %  left, bottom, right and top
\includegraphics[trim=0.0cm 0.0cm 0.0cm 0.0cm, clip=false, width=\textwidth]{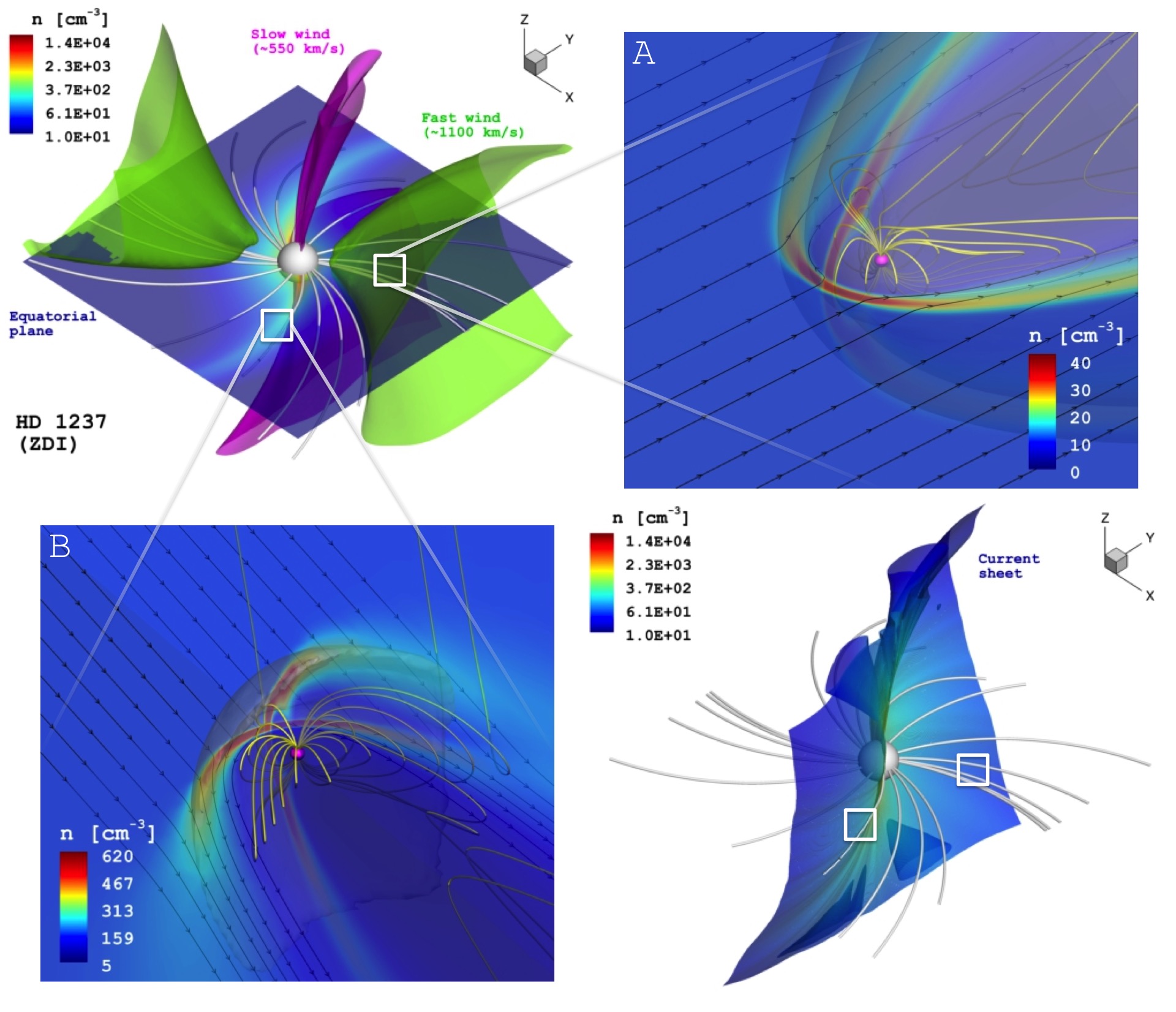}
\caption{Simulated environment of the HD 1237 system driven by the ZDI magnetic field map. The structure of the stellar wind and astrospheric current sheet obtained from the IH module, are presented in the top-left and bottom-right panels respectively. The density structure of the steady-state solution is displayed on the equatorial plane (top-left) and the astrospheric current sheet (bottom-right). In the top-left panel, the topology and associated magnitudes of the dominant radial velocity components ($u_{\rm r}$) of the stellar wind are also included (fast: green -- slow: magenta). The central white sphere denotes the boundary with the SC domain at 25\,$R_{*}$ (Sect. \ref{sec_alfven}), and selected 3D stellar wind magnetic field lines are shown in white. The two remaining panels contain the simulation results of the GM module, obtained at the locations indicated on the IH domain by the white squares (not to scale). The distance to the star has been taken as the mean orbital separation of this system ($a = 0.49$ AU, \citeads{2001A&A...375..205N}). The central purple sphere corresponds to the planetary surface (1 $R_{p}$) and selected 3D planetary magnetic field lines are displayed in yellow. The direction of the incident stellar wind is indicated by the black streamlines. The particle density distribution of the solution shows the development of a bow-shock structure in both cases (translucent white shade).}
\label{fig_8}
\end{figure*}

\begin{figure*}[ht]
\centering %  left, bottom, right and top
\includegraphics[trim=0.0cm 0.0cm 0.0cm 0.0cm, clip=false, width=\textwidth]{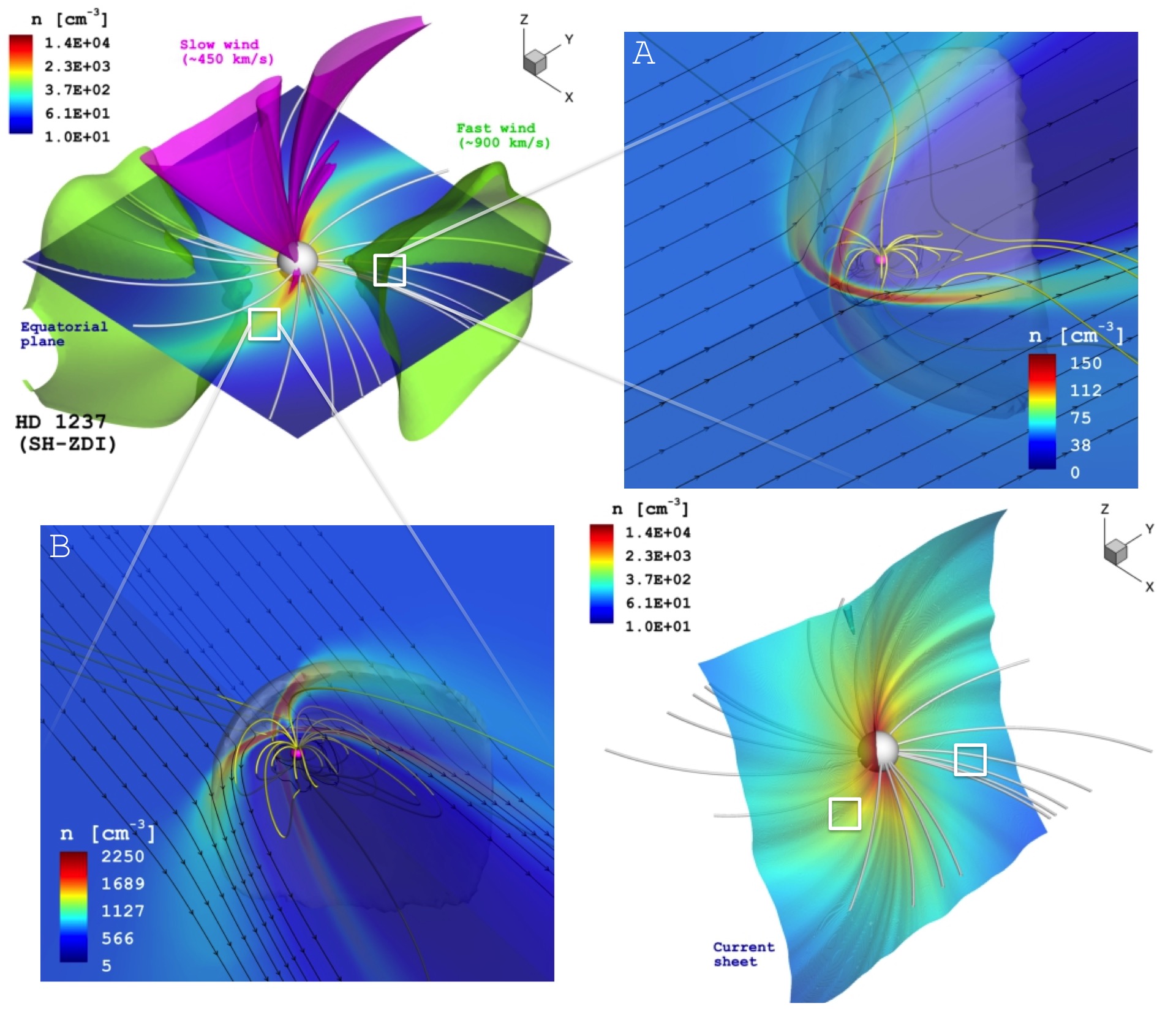}
\caption{Simulated environment of the HD 1237 system driven by the SH-ZDI magnetic field map. The structure of the stellar wind and astrospheric current sheet obtained from the IH module, are presented in the top-left and bottom-right panels respectively. The density structure of the steady-state solution is displayed on the equatorial plane (top-left) and the astrospheric current sheet (bottom-right). In the top-left panel, the topology and associated magnitudes of the dominant radial velocity components ($u_{\rm r}$) of the stellar wind are also included (fast: green -- slow: magenta). The central white sphere denotes the boundary with the SC domain at 25\,$R_{*}$ (Sect. \ref{sec_alfven}), and selected 3D stellar wind magnetic field lines are shown in white. The two remaining panels contain the simulation results of the GM module, obtained at the locations indicated on the IH domain by the white squares (not to scale). The distance to the star has been taken as the mean orbital separation of this system ($a = 0.49$ AU, \citeads{2001A&A...375..205N}). The central purple sphere corresponds to the planetary surface (1 $R_{p}$) and selected 3D planetary magnetic field lines are displayed in yellow. The direction of the incident stellar wind is indicated by the black streamlines. The particle density distribution of the solution shows the development of a bow-shock structure in both cases (translucent white shade).}
\label{fig_9}
\end{figure*}

\noindent The results of the simulations performed on the HD 1237 system, driven by the ZDI and SH-ZDI maps, are presented in Figs. \ref{fig_8} and \ref{fig_9} respectively. Two fast wind structures oriented in opposite directions, appear close to the equatorial region of the system. These structures are connected to the large, low-latitude coronal holes developed in this system (see \citeads{2016A&A...588A..28A}). The wind speed reaches $\sim$1100 km s$^{-1}$ in the ZDI-driven case, dropping to $\sim$900 km s$^{-1}$ in the SH-ZDI simulation. As  in some of the previously described cases, the fast wind regions appear roughly perpendicular to the astrospheric current sheet, along which the slow wind region develops. The global topology of the stellar wind is similar between both cases, yet an enhancement in the particle density is again obtained in the SH-ZDI case (see Sects. \ref{sec_alfven} and \ref{sec_resultsIH}). A more quantitative comparison of all cases is presented in the following section.

As described in Sect. \ref{sec_model}, the GM module of the SWMF was additionally coupled to the IH solution to investigate the exoplanetary conditions in relation to the developed stellar wind properties in this system. Two different locations in the IH domain, represented by the white squares in Figs. \ref{fig_8} and \ref{fig_9}, were used for this purpose. These locations correspond to a fast wind, low density region (sector A | Figs. \ref{fig_8} and \ref{fig_9}, top-right), and a high-density streamer of the stellar wind, close to the astrospheric current sheet (sector B | Figs. \ref{fig_8} and \ref{fig_9}, bottom-left). The distance to the central star was $0.49$ AU in both cases, matching the mean orbital separation of this system (\citeads{2001A&A...375..205N}). The interaction between the magnetised stellar wind and the planetary magnetosphere, led to the development of a bow-shock structure, which self-consistently reacts to the local conditions. A summary of the driving stellar wind properties, and the resulting magnetospheric conditions in both locations, is presented in Table \ref{table_2}. 

\noindent A larger response from the magnetosphere was obtained in the SH-ZDI case as expected, given the incident stellar wind properties in the IH domain (see Table \ref{table_2}). In the fast wind region (location A in Figs. \ref{fig_8} and \ref{fig_9}), the maximum total pressure at the bow-shock ($P^{\rm\,max}_{\rm T}$) appeared $\sim$\,2 times larger in the SH-ZDI case compared to the ZDI simulation. This led to $\sim$\,15\% reduction in the magnetosphere size (in terms of the magnetopause standoff distance $R_{\rm M}$) in the former case compared to the latter. A similar situation was obtained in the simulations at the high-density streamer sector (location B in Figs. \ref{fig_8} and \ref{fig_9}). The SH-ZDI case yielded a $\sim$\,3.7 times larger $P^{\rm\,max}_{\rm T}$ value, and a $\sim$\,25\% reduction in $R_{\rm M}$, in comparison with the ZDI-driven model. On the other hand, the resulting conditions were more extreme in the dense streamer sector (location B), than in the fast wind region (location A), regardless of the driving field distribution. This is evidenced by the $\sim$\,10 times increase in the average particle density $\left<n\right>$, between locations A and B (see Table \ref{table_2}), calculated at $1 R_{p}$ above the planetary surface (i.e. over a spherical surface with $R = 2 R_{p}$). As can be seen for Figs. \ref{fig_8} and \ref{fig_9}, the developed wind structure shows a larger contrast in density than in velocity. This general property was common in the remaining systems considered here and in the solar cases (see Sect. \ref{sec_resultsIH}). 

\begin{table*}[!htbp]
\caption{Parameters of the incident stellar wind and resulting properties inside the GM module for each location.}             
\label{table_2}      
\centering
{\small        
\begin{tabular}{ l | c c c c | c c c }    
\hline\hline
\\[-9pt]
Location / &  \multicolumn{4}{|c|}{Incident stellar wind (IH module) \tablefootmark{$\dagger$}} & \multicolumn{3}{|c}{Global Magnetosphere (GM module)} \\
Case & $n$ [cm$^{-3}$] & $T$ [$\times$\,10$^{5}$ K] & $\mathbf{u_{\rm sw}}$ [km s$^{-1}$] & $\mathbf{B}$ [nT] & $R_{\rm M}$ [$R_{\rm p}$] \tablefootmark{a} & $P^{\rm\,max}_{\rm T}$ [nPa] \tablefootmark{b} & $\left<n\right>$ [cm$^{-3}$] \tablefootmark{c}\\[2pt]
\hline
& & & & & & &\\[-9pt] 
A / ZDI & 10.9 & 5.37 & ($-829.5, 804.8, -5.4$) & ($-16.5, 10.5, 0.01$) & 7.7 & 51.5 & 1.13 \\
A / SH-ZDI & 45.5 & 3.40 & ($-641.3, 638.2, -0.2$) & ($-43.7, -28.7, 0.31$) & 6.5 & 101.7 & 3.05 \\
B / ZDI & 160.0 & 1.03 & ($-383.7, 466.4, -6.3$) & ($1.22,-3.31,-0.18$) & 6.4 & 165.2 & 9.13 \\
B / SH-ZDI & 596.7 & 1.12 & ($-222.5, 553.7, -5.8$) & ($1.15, -23.1, 0.60$) & 4.7 & 618.7 & 35.1 \\[1pt]
\hline                  
\end{tabular}}
\tablefoot{\tablefoottext{a}{This value corresponds to the magnetopause standoff distance (i.e. magnetopause day-side separation).}\tablefoottext{b}{Maximum value of the total pressure ($P_{\rm T} = P_{\rm gas} + P_{\rm dyn} + P_{\rm mag}$) at the bow-shock location.}\tablefoottext{c}{Average particle density at a spherical surface with $R = 2 R_{p}$.}\\ \tablefoottext{$\dagger$}{Cartesian components of the vector quantities are provided.}} 
\end{table*}

\noindent In this context, the results obtained for locations A and B indicate that the process of particle injection into the planetary atmosphere, would be more sensitive to the density structure rather than to the velocity profile of the stellar wind. 

Finally, and linked to this last result, we can consider the simulated wind structure and the exoplanetary orbit to investigate the possible magnetospheric radio emission from the exoplanet of this system. The theoretical considerations for this kind of emission have been presented in various papers (e.g. \citeads{1999JGR...10414025F}; \citeads{2001Ap&SS.277..293Z}; \citeads{2007A&A...475..359G}; \citeads{2008A&A...490..843J}; \citeads{2011MNRAS.414.2125N}; \citeads{2016arXiv160603997N}), and extensive observational searches have been performed over the last decade (e.g. \citeads{2007ApJ...668.1182L}; Lazio et~al. \citeyearads{2010AJ....139...96L}, \citeyearads{2010AJ....140.1929L}; \citeads{2007MNRAS.382..455G}; Lecavelier et~al. \citeyearads{2009A&A...500L..51L}, \citeyearads{2011A&A...533A..50L}, \citeyearads{2013A&A...552A..65L}; \citeads{2013ApJ...762...34H}; \citeads{2014A&A...562A.108S}). For the particular case of HD\,1237, \citetads{2005MNRAS.356.1053S} identified this system as a good  candidate for detection, with a mean radio flux of 8 mJy at a peak frequency of $\sim$\,40 MHz, reaching up to $\sim$\,20 mJy during periastron passage. Lacking better information, \citetads{2005MNRAS.356.1053S} used reasonable approximations regarding the stellar wind and planetary properties, such as a scaling relation from \citetads{2002ApJ...574..412W} between the X-ray flux and the mass loss rate of the host star (see Sect. \ref{sec_B-Mloss}), spherical symmetry for the stellar wind, scaling of the planetary magnetic moment and radius, among others. While no better constraints are available for the planetary properties, our data-driven simulations provide a more realistic description of the wind of this star. For this reason, we retained the assumptions made by \citetads{2005MNRAS.356.1053S} regarding the exoplanet properties of radius and magnetic moment $\mathcal{M}_{\rm p}$ (i.e., $R_{\rm p} = R_{\tiny\jupiter} \simeq 7.2\,\times\,10^{9}$ cm; $\mathcal{M}_{\rm p} = \mathcal{M}_{\tiny\jupiter} \simeq 1.6\,\times\,10^{30}$ G cm$^3$). This implies that the 40 MHz peak frequency of the expected magnetospheric radio emission remains unaltered (see \citeads{2005MNRAS.356.1053S}). For this analysis we only consider the results from the SH-ZDI simulation, as it provides favourable conditions in terms of increased stellar wind density (See Figs. \ref{fig_8} and \ref{fig_9}). 

Following \citetads{2001Ap&SS.277..293Z}, the emitted radio power from the exoplanet, $R^{\rm pow}$, will be proportional to the kinetic power $K_{\rm sw}^{\rm pow}$ associated with the wind-magnetosphere interaction (i.e., $R^{\rm pow} = \alpha K_{\rm sw}^{\rm pow}$, with $\alpha = 7\,\times10^{-6}$\footnote[2]{This relation is known as the \textit{Radiometric Bode Law}. See \citetads{2001Ap&SS.277..293Z} and \citetads{2004ApJ...612..511L}}). By combining the simulated stellar wind structure with the assumed exoplanetary properties, we can compute $K_{\rm sw}^{\rm pow}$, using the relation 

\begin{equation}\label{eq_1}
K_{\rm sw}^{\rm pow} = nu_{\rm sw}^3\pi R_{\rm M}^2\mbox{ ,}
\end{equation}

\noindent where $R_{\rm M}$ denotes radius of the magnetosphere, which depends on the local conditions of the wind and the planetary magnetic field (e.g. Table \ref{table_2}). As discussed by \citetads{2005MNRAS.356.1053S} and references therein, $R_{\rm M}$ can be expressed as 

\begin{equation}\label{eq_2}
R_{\rm M} \propto \left(\dfrac{\mathcal{M}_{\rm p}^2}{16\pi nu_{\rm sw}^2}\right)^{1/6}\mbox{ ,} 
\end{equation}

\noindent and therefore, can be calculated at each point of the simulation domain. However, as the exoplanet location is not arbitrary, only values of $K_{\rm sw}^{\rm pow}$ along the planetary orbit will be relevant for the predicted magnetospheric radio emission. 

\begin{figure}[ht]
\centering %  left, bottom, right and top
\includegraphics[trim=0.0cm 0.0cm 0.0cm 0.0cm, clip=true,width=\hsize]{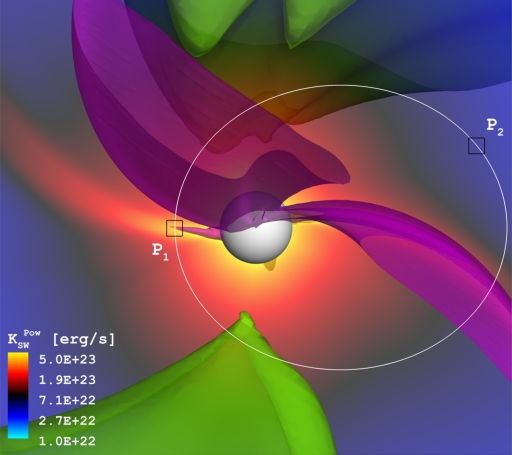}\caption{Pole-on view on the equatorial distribution of the kinetic power $K_{\rm sw}^{\rm pow}$ from wind-magnetospheric interaction (see text for details). The green (fast) and magenta (slow) velocity components of the stellar wind are identical as in Fig. \ref{fig_9} (top-left). The configuration of the exoplanet orbit (white ellipse), with respect to the structure of the stellar wind, maximises the radio power at P$_1$ (periastron), and yields a minimum in location P$_2$.}
\label{fig_10}
\end{figure}

\noindent Figure \ref{fig_10} shows the equatorial distribution of $K_{\rm sw}^{\rm pow}$, alongside the dominant radial wind components of the SH-ZDI simulation of this system (pole-on view of Fig. \ref{fig_9}, top-left). The exoplanet orbit, indicated by the white ellipse, has been constructed using the parameters listed by \citetads{2001A&A...375..205N}. The wind-orbit layout presented in Fig. \ref{fig_10} corresponds to the optimal conditions for magnetospheric radio emission in this system (i.e., periastron passage through a dense wind streamer, location P$_{1}$ in Fig. \ref{fig_10}). In this way, we obtain a maximum emitted radio power of $R^{\rm pow} = \alpha K_{\rm sw}^{\rm pow} \simeq 3.5\,\times10^{18}$ erg s$^{-1}$, which yields a corresponding radio flux on Earth\footnote[2]{We have used here the distance to HD 1237 of 17.5 pc (\citeads{2010MNRAS.403.1949K}).} of $F_{\oplus}^{R} \simeq 12$ mJy. This value is reduced up to a factor of $\sim$\,10 in the location marked as P$_2$ in Fig. \ref{fig_10}, which does not coincide with apastron in this configuration. 

Our maximum radio flux $F_{\oplus}^{R}$ is roughly half of the previous estimate of 21.5 mJy of \citetads{2005MNRAS.356.1053S}. In turn, the predicted orbital variation in our simulation is more than twice as large as the $\sim$\,4.5 factor obtained in this previous work. Given that this previous analysis was performed assuming a stellar wind velocity of 400 km s$^{-1}$ (slower than the value predicted in our simulations at periastron by $\sim$\,25\%), the higher radio power obtained by \citetads{2005MNRAS.356.1053S} must be connected with the assumed stellar wind density (obtained via a spherically symmetric wind with $\dot{M} = 85.7 \dot{M}_{\odot}$). As will be discussed in the following section, this mass loss rate value is probably overestimated, supporting a lower value of the planetary radio flux. The difference in the orbital variation of $F_{\oplus}^{R}$ appears as a consequence of a more realistic description of the 3D stellar wind structure provided by our data-driven simulation. 

Finally, the obtained values for the magnetospheric radio emission should be within the expected capabilities of the Square Kilometre Array (SKA), particularly in the low-frequency band (50 MHz up to 350 MHz, see \citeads{2015aska.confE.120Z}). The on-sky location of this system [$\alpha$ (J2000): $00^{\rm h} 16^{\rm m} 16.68^{\rm s}$, $\delta$ (J2000): $-79^\circ 51\arcmin 04.25\arcsec$], prevents observations with current instrumentation with low-frequency capabilities, such as the Low-Frequency Array (LOFAR) and the Ukrainian T-shaped Radio telescope (UTR-2). 

\section{Analysis and discussion}\label{sec_Analysis}

In a similar manner to \citetads{2016A&A...588A..28A}, we use the simulation results to analyse several aspects of the environment of these systems. In this way, we consider the connection between the surface magnetic field properties, and the predicted mass and angular momentum loss rates associated with the wind (Sect. \ref{sec_B-Mloss}). In addition, a characterisation of the stellar wind properties at the inner edges of the habitable zones of these systems is presented in Sect. \ref{sec_wind-HZ}. 

\subsection{Magnetism and mass/angular momentum loss rates}\label{sec_B-Mloss}

\begin{figure}[!t]
\centering %  left, bottom, right and top
\includegraphics[trim=0.2cm 0.0cm 0.5cm 0.0cm, clip=true,width=\hsize]{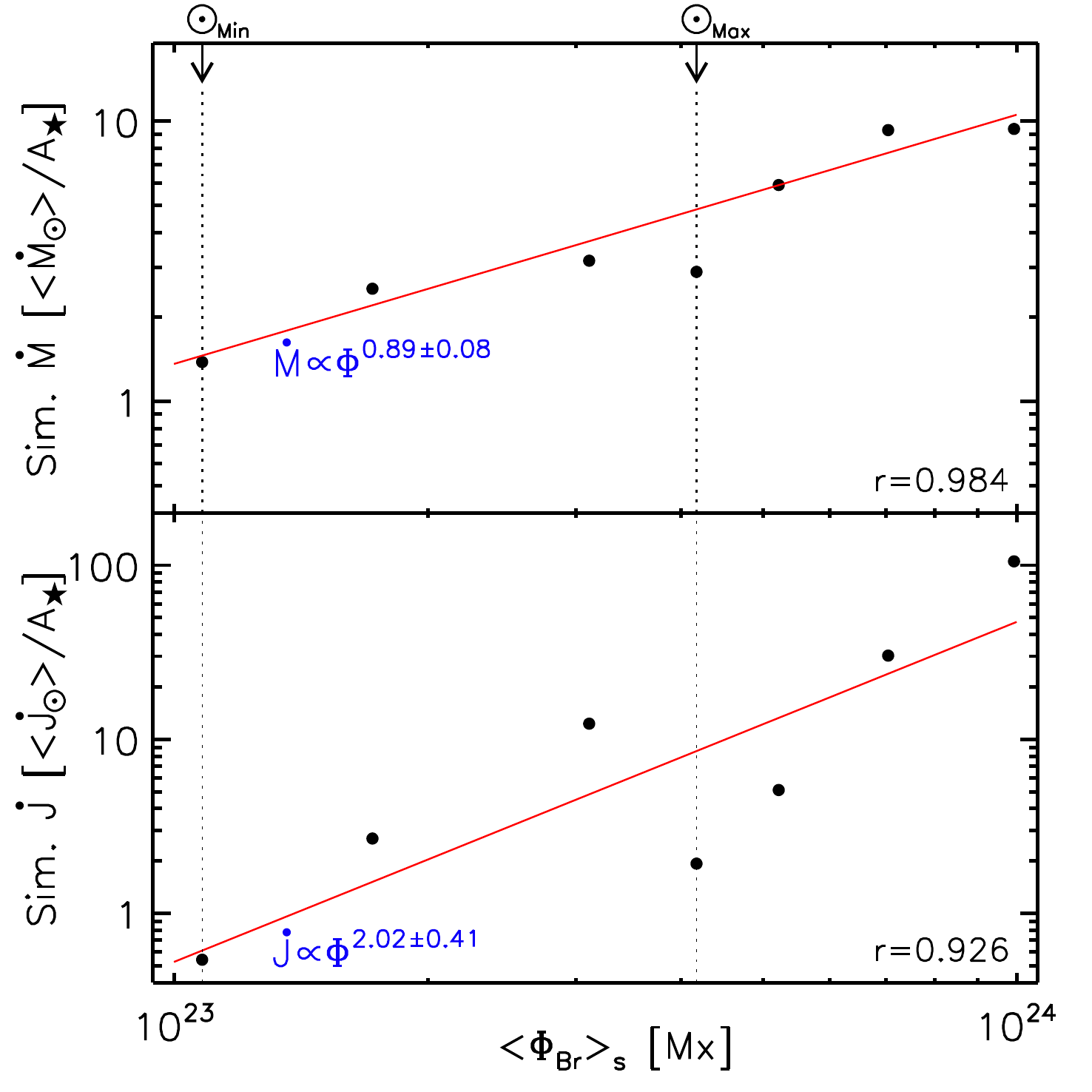}
\caption{Simulated mass loss rate ($\dot{M}$, top) and angular momentum loss rate ($\dot{J}$, bottom) as a function of the surface-averaged unsigned radial magnetic flux $\left<\Phi_{\rm Br}\right>_{\rm s}$. Both quantities are expressed in units of average solar values (Sect. \ref{sec_alfven}), normalised by the surface area of each star. Individual points denote the results of each simulation presented in Sect. \ref{sec_results}, including the solar cases (indicated by the dashed vertical lines). The red solid lines show power-law fits to the simulated data with the corresponding correlation coefficient, $r$, in each case.}
\label{fig_11}
\end{figure}

As with the first paper of this study, this analysis considers the results of our simulations independently (i.e. solar min/max and stellar ZDI/SH-ZDI cases). Figure \ref{fig_11} shows the dependence of the simulated mass and angular momentum loss rates ($\dot{M}$ and $\dot{J}$, respectively), with respect to the unsigned radial magnetic flux $\Phi_{\rm Br}$, averaged over the stellar surface. An approximately linear dependence is obtained for $\dot{M}$ ($\propto\Phi^{0.89\,\pm\,0.08}$)\footnote[3]{For simplicity, in the following relations $\Phi$ represents $\left<\Phi_{\rm Br}\right>_{\rm s}$.}, while a quadratic relation (with increased scatter) is obtained for $\dot{J}$ ($\propto\Phi^{2.02\,\pm\,0.41}$). Given the direct dependence of $\dot{J}$ with $\dot{M}$ (see \citeads{2010ApJ...721...80C}; \citeads{2012ApJ...754L..26M}; \citeads{2014ApJ...783...55C}; \citeads{2015ApJ...798..116R}; \citeads{2015ApJ...799L..23M}; \citeads{2015ApJ...813...40G}), we will focus our discussion on the results obtained for the mass loss rate. A diagram similar to Fig \ref{fig_11}, relating the simulated coronal radiation (e.g. EUV, X-rays) with $\left<\Phi_{\rm Br}\right>_{\rm s}$ was presented in \citetads{2016A&A...588A..28A}. 

Unlike the high-energy emission, mass loss rate estimates are only available for 10 Sun-like stars (spectral types G-K). This sample covers activity levels expressed in terms of X-ray fluxes ($F_{\rm X}$) between $10^4$ to few 10$^6$ erg cm$^{-2}$ s$^{-1}$, and includes 5 multiple systems ($\alpha$ Cen, 61 Cyg, $\xi$ Boo, 36 Oph and 70 Oph), and 5 single stars (61 Vir, Sun, $\epsilon$ Ind, $\epsilon$ Eri and $\pi^1$ UMa). As mentioned in Sect. \ref{sec_intro}, these indirect measurements are obtained via the hydrogen wall in the astrosphere of the system, which is detected as extra H\,I Lyman-$\alpha$ absorption in the UV region of the spectra (Wood et al. \citeyearads{2005ApJ...628L.143W}, \citeyearads{2014ApJ...781L..33W}). In the low and moderate activity regime ($F_{\rm X} < 10^6$ erg cm$^{-2}$ s$^{-1}$), a mass loss-activity relation in the form of $\dot{M} \propto F^{1.34 \pm 0.18}_{\rm X}$ was suggested by \citetads{2005ApJ...628L.143W}. This relation appears to break for the high-activity end ($F_{\rm X} > 10^6$ erg cm$^{-2}$ s$^{-1}$), where $\xi$ Boo A and $\pi^1$ UMa have been recently located with mass loss estimates $\dot{M} < \dot{M}_{\odot}$ for both stars (\citeads{2014ApJ...781L..33W})\footnote[4]{Two very active M dwarfs, Proxima Cen (upper limit) and EV Lac, are also located in this region, with small (absolute) mass loss rate values ($\dot{M} \le \dot{M}_{\odot}$). Their location in \citetads{2014ApJ...781L..33W} diagram is due to their very small surface areas ($\sim$\,$0.023$ A$_{\odot}$ and $0.123$ A$_{\odot}$, respectively).}. 

We can relate these observational studies with our simulations, by combining the $\dot{M}$\,--\,$\Phi$ relation shown in Fig. \ref{fig_11}, with the $L_{\rm X} \propto \Phi^{1.06}$ dependence obtained in \citetads{2016A&A...588A..28A}. The latter becomes slightly more steep when expressed in terms of $F_{\rm X}$ (i.e. $F_{\rm X} \propto \Phi^{1.13}$). By removing the dependence on $\Phi$, we obtain a simulated mass loss-activity relation in the form of $\dot{M}_{\rm (sim)} \propto F_{\rm X\,(sim)}^{\rm \,\gamma}$ with $\gamma = 0.79^{\,+\,0.19}_{\,-\,0.15}$, considerably flatter than the observed one. However, we stress that this is the first time that this relation is self-consistently constructed in a model, by computing the X-ray coronal emission and mass loss rate associated with the stellar wind, using the same data-driven numerical simulation. To understand the differences in these relations, we need to consider several aspects connected to our simulations and the observations. A comprehensive summary is listed below:

\begin{itemize}
\item [-] As presented in Sect. \ref{sec_results}, the results from the solar simulations are consistent with the observational data. This is not only the case for the mass loss rate (Sect. \ref{sec_alfven}) but also for the topology and physical properties of the solar wind during activity minimum and maximum (Sect. \ref{sec_resultsIH}, Fig. \ref{fig_5}). Similar results were obtained for the simulated coronal structure for both activity states (\citeads{2016A&A...588A..28A}).

\smallskip

\item[-] For the stellar systems considered here, an observational estimate of the mass loss rate is only available for HD\,22040 ($\epsilon$ Eri, \citeads{2002ApJ...574..412W}). As can be seen from Table \ref{table_1}, the ZDI-driven simulation predicts an absolute mass loss rate $\dot{M}_{\rm ZDI} \simeq \dot{M}_{\odot}$, while the SH-ZDI case leads to $\dot{M}_{\rm SH-ZDI} \sim 5\,\dot{M}_{\odot}$\footnote[2]{The reasons for this relative difference are discussed in Sect. \ref{sec_alfven}.}. This last value differs by a factor of 6 from the estimation from astrospheric Lyman-$\alpha$ absorption ($\dot{M}_{\rm Ly-\alpha} \simeq 30\,\dot{M}_{\odot}$). 

\smallskip

For the other two stars, HD\,1237 and HD\,147513, \citetads{2005MNRAS.356.1053S} derived relatively high $\dot{M}$ values ($\sim$\,86 $\dot{M}_{\odot}$ and 105 $\dot{M}_{\odot}$, respectively), by using an earlier version of the mass loss-activity relation proposed in \citeads{2002ApJ...574..412W} (i.e. $\dot{M} \propto F_{\rm X}^{1.15 \pm 0.20}$). As described before, the range of validity of this relation was revisited by \citetads{2005ApJ...628L.143W} and \citetads{2014ApJ...781L..33W}, indicating a break for $F_{\rm X} > 10^{6}$ erg cm$^2$ s$^{-1}$, with evidence in support of much smaller $\dot{M}$ values for stars in this activity regime. This evidence is not only given directly by the astrospheric detections of $\xi$ Boo A and $\pi^1$ UMa (\citeads{2014ApJ...781L..33W}), but also indirectly, by similar UV observations of a considerable number of active stars yielding non-detections (\citeads{2005ApJS..159..118W}). Both  stars, HD\,1237 and HD\,147513, have X-ray fluxes above this empirical $F_{\rm X}$ threshold (by factors of 2.1 and 1.9, respectively), thus the values listed in \citetads{2005MNRAS.356.1053S} are probably overestimated.

\smallskip

\item[-] In the case of HD\,22049, this discrepancy may be addressed by enhancing the base conditions in the simulation of this star. While this certainly would increase the mass loss rate (see \citeads{2014ApJ...783...55C}), this would also imply a different prediction for the coronal emission in all energy bands (e.g. EUV, SXR). As discussed in \citetads{2016A&A...588A..28A}, the SH-ZDI simulation of this system provides reasonable agreement in both, EUV and X-rays, to the estimated coronal conditions via spectral synthesis diagnostics (\citeads{2011A&A...532A...6S}). Still, further adjustments will be explored for this system in a future systematic approach, in order to improve the balance in the coronal heating (i.e. to the Emission Measure distribution $EM$, see \citeads{2008MNRAS.385.1691N}; \citeads{2016A&A...588A..28A}), as well as to refine the predictions of the $\dot{M}$ value. It might even be necessary to include reconnection events as an additional (or dominant) heating mechanism, which appear to drive the coronal conditions in very active stars such as HD\,22049 (see \citeads{2000ApJ...545.1074D}). As a drawback, this procedure introduces additional degrees of freedom for the model results, which then complicates the consistent comparison with the reference the solar cases (and with additional stellar simulations).  

\smallskip

\item[-] Another possibility is related to a temporal dependence of the mass loss-activity relation. In the case of the Sun, it is well known that the coronal emission is enhanced by one order of magnitude over the course of the 11-year activity cycle (\citeads{lrsp-2015-4}). In turn the solar mass loss rate shows little correspondence with the activity state, fluctuating within a factor of $\sim$\,2 around a mean value of $\dot{M}_{\odot}$\,$\simeq$\,$2\times\,10^{-14}$ $M_{\odot}$ yr$^{-1}$ (Sect. \ref{sec_alfven}, see also \citeads{2011MNRAS.417.2592C}). While there is at least one single star showing a similar cycle-induced\footnote[3]{The cycle length ($\sim\,$1.6 years) is much shorter than the solar cycle, as expected from the relatively young age of the star ($\sim\,$500 - 740 Myr, see \citeads{2013A&A...553L...6S}).} pattern as the Sun in X-rays ($\iota$ Horologii, see \citeads{2013A&A...553L...6S}), there is no evidence suggesting that stellar mass loss rates must be cycle-independent (or as near-to-independent as the Sun is). Therefore, variations of 1\,-\,2 orders of magnitude in $\dot{M}$, over the course of any possible activity cycle (or due to cycle-dominated transients such as Coronal Mass Ejections - CMEs, see \citeads{2013ApJ...764..170D}), cannot be excluded from the mass loss-activity relation. This would provide a natural explanation for the observed break in this relation, at the regime of high coronal activity.   

\smallskip

\item[-] Given the magnetic nature of these two processes in Sun-like stars (coronal activity and mass loss), any time dependence of the mass loss-activity relation should be connected with the temporal evolution of the stellar magnetic field. Unfortunately, for the two stars considered in this work located above the break in this relation (HD\,1237 and HD\,147513), only single-epoch\footnote[4]{Actually, two independent ZDI maps were recovered for HD\,1237 separated by 5 months. However, the quality of the second map was far from optimal, due to a very limited phase coverage (see \citeads{2015A&A...582A..38A}). Still, this partial field reconstruction indicated very similar properties as with the robust ZDI maps of the first epoch, which have been used to drive the models presented here.} surface magnetic field reconstructions using ZDI are available (\citeads{2015A&A...582A..38A}; \citeads{2016A&A...585A..77H}). 

\smallskip On the other hand, a long-term ZDI monitoring campaign of HD\,22049 is currently being carried out by the BCool\footnote[1]{\url{http://bcool.ast.obs-mip.fr}} collaboration. Six large-scale magnetic field maps have been recovered over a period of 7 years (2007\,--\,2013, see \citeads{2014A&A...569A..79J}). As described in the first paper of this study, we have used the set of HARPSpol@ESO3.6m observations available for this star so far (acquired in 2010, \citeads{2011Msngr.143....7P}), to generate the ZDI maps driving the simulations. This was done to ensure a consistent comparison with ZDI-driven models of the other two stars (whose maps were recovered also using HARPSpol), by applying the same procedures and criteria in the reconstructions (see Alvarado-G\'omez et al. \citeyearads{2015A&A...582A..38A}, \citeyearads{2016A&A...588A..28A}), minimising at the same time the effects introduced by driving the simulations using maps from different instruments. By relaxing these last requirements, we can make an order-of-magnitude estimate of the possible temporal variations in $\dot{M}$ and $F_{\rm X}$ of HD\,22049, due to the long-term evolution of its large-scale magnetic field. For this we use the ZDI information provided by \citetads{2014A&A...569A..79J}, together with the $F_{\rm X}$\,--\,$\Phi$ dependancy derived from the results in \citetads{2016A&A...588A..28A}, and the $\dot{M}$\,--\,$\Phi$ relation presented in Fig. \ref{fig_11}. 

\smallskip Following this procedure, the recovered large-scale field distributions indicate an approximate change in $\left<\Phi_{\rm Br}\right>_{\rm s}$ by a factor of $\sim$2 in a time-scale of at least 5 years\footnote[2]{This is obtained by considering the largest difference (including uncertainties) in either, the maximum ($B_{\rm max}$) or the mean ($B_{\rm mean}$) magnetic field values listed by \citetads{2014A&A...569A..79J}. For the former, this occurred between the epochs of 2007 and 2012, while for the latter this was visible among the maps of 2008 and 2013.}. This would imply variations in $F_{\rm X}$ and $\dot{M}$ up to factors of $\sim$\,2.4 and 2.1, respectively. We stress here that this calculation corresponds to a first-order approximation, as we are neglecting several important elements such as the field topology, complexity, missing magnetic flux, and map incompleteness, among others, which are known to influence the predictions of $F_{\rm X}$ and $\dot{M}$ based on ZDI maps (e.g. \citeads{2011MNRAS.410.2472A}; \citeads{2013ApJ...764...32G}; \citeads{2014MNRAS.439.2122L}; \citeads{2015ApJ...807L...6G}; \citeads{2015ApJ...813...40G}; \citeads{2016A&A...588A..28A}). Nevertheless, archival X-ray observations of this star, available at the Nearby X-ray and Extreme UV Emitting Stars (NEXXUS 2) database\footnote[3]{\url{http://www.hs.uni-hamburg.de/DE/For/Gal/Xgroup/nexxus/nexxus.html}} (\citeads{2004A&A...417..651S}), indicate a variation in $F_{\rm X}$ within a time-scale of 10 years compatible with our previous estimate. Such change is sufficient to move HD\,22049 above the previously mentioned threshold in the mass-loss activity relation ($F_{\rm X} = 10^6$ erg cm$^{-2}$ s$^{-1}$). Assuming a physical origin for this apparent break, the variation in $F_{\rm X}$ would imply a decrease in the mass loss-rate value of HD\,22049, much larger than the one estimated above and reaching similar values as the ones predicted by our simulations. 

\smallskip

\item[-] Last but not least, we must also consider the uncertainties associated with the observational estimates of mass loss rates. As explained by \citetads{2005ApJ...628L.143W} and \citetads{2014ApJ...781L..33W}, robust astrospheric detections require precise knowledge of the physical structure of the local ISM (e.g. the case of $\lambda$ And, see \citeads{2014ApJ...787...75M}). This includes column densities, kinematics, and metal depletion rates (\citeads{2004ApJ...602..776R}; \citeads{2008ApJ...683..207R}), together with local temperature and turbulent velocities (\citeads{2004ApJ...613.1004R}), interpreted within a particular model of the morphology of the ISM (see Redfield \& Linsky \citeyearads{2008ApJ...673..283R}, \citeyearads{2015ApJ...812..125R}; \citeads{2014A&A...567A..58G}). While these studies have provided a detailed characterisation of the local ISM, intrinsic uncertainties and additional observational issues connected with the astrospheric detections, can certainly modify the estimated mass loss rates by large amounts (conservatively, by factors $\sim\,$2\,--\,3; see \citeads{2014ASTRP...1...43L}; \citeads{2002ApJ...574..412W}).  
\end{itemize}

\noindent Some of the possibilities discussed before are currently being explored and will be presented in a future paper. This study will follow the same data-driven methodology presented here, applied to 70\% of the stars with astrospheric detections (\citeads{2014ApJ...781L..33W}) and with surface magnetic fields distributions recovered by ZDI\footnote[4]{Additional ZDI maps have been recovered using the NARVAL spectro-polarimeter (\citeads{2003EAS.....9..105A}), from observations acquired during 2015 (Program ID: L151N08 -- PI: Morin).} (see \citeads{2016MNRAS.455L..52V} and references therein). 

\subsection{Stellar winds and habitable zones}\label{sec_wind-HZ}

\noindent Finally, we can use our 3D simulation results, to characterise the stellar wind conditions in the estimated boundaries of the Habitable Zones (HZ) of these systems. For the latter, we take advantage of the information provided in the Habitable Zone Gallery\footnote[1]{\url{http://www.hzgallery.org/}} (HZG, \citeads{2012PASP..124..323K}), regarding the \textit{optimistic} and \textit{conservative} calculations of the HZ (as defined in Kopparapu et al. \citeyearads{2013ApJ...765..131K}, \citeyearads{2014ApJ...787L..29K}). We restrict our analysis to the inner edges of the HZs, given the enhancement of various stellar wind properties (e.g. density, magnetic field) closer to the star, which could influence the very definition of this boundary. Reference calculations at 1 AU are also performed for all the simulated cases. In the case of the Sun, the inner edges of the HZ are extremely close to 1 AU, thus we consider instead the semi-major axes of Mercury ($a_{\tiny\mercury}$\,$\simeq$\,0.39\,AU) and Venus ($a_{\tiny\venus}$\,$\simeq$\,0.72\,AU).   

\begin{figure}[ht]
\centering %  left, bottom, right and top
\includegraphics[trim=0.0cm 0.1cm 0.8cm 0.1cm, clip=true,width=\hsize]{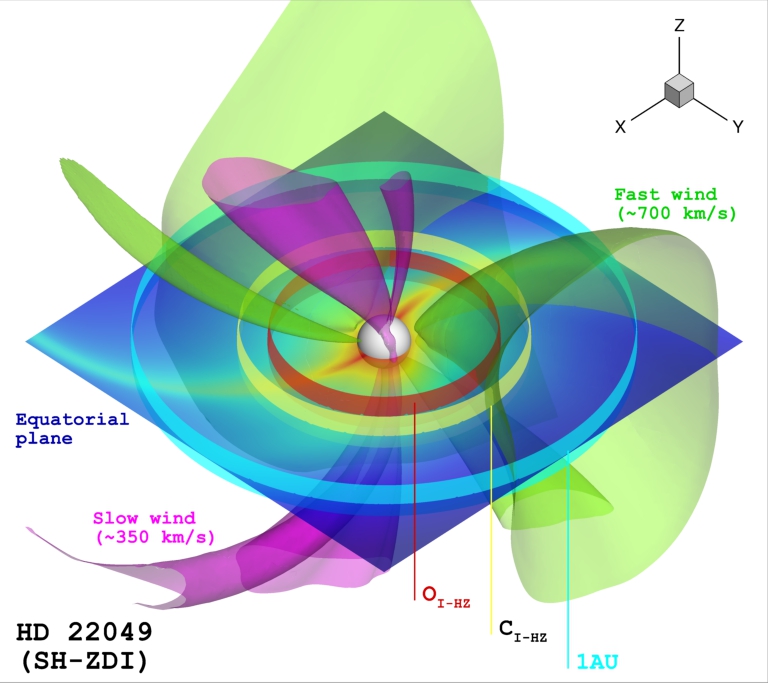}
\caption{Stellar wind characterisation at the optimistic (O$_{\rm I-HZ}$, red) and conservative (C$_{\rm I-HZ}$, yellow) inner edges of the HZ of HD\,22049. Similarly, the 1 AU boundary is used as reference (cyan). In all cases, we consider rings extending to $\pm\,5^{\circ}$ in inclination from the equatorial plane (slightly exaggerated in the figure for visualisation purposes). The wind structure is the same as in Fig. \ref{fig_6} (bottom-left), rotated by 90$^{\circ}$ clockwise.}
\label{fig_12}
\end{figure}

\noindent  As an example, Fig. \ref{fig_12} shows the optimistic (O$_{\rm I-HZ}$, red) and conservative (C$_{\rm I-HZ}$, yellow) inner edges of the HZ of HD\,22049. These are displayed alongside the 1 AU boundary (cyan), and the stellar wind structure developed in the IH domain of the SH-ZDI simulation (Sect. \ref{sec_resultsIH}, Fig. \ref{fig_6}). As presented in the visualisation, we consider rings extending $\pm\,5^{\circ}$ in inclination from the equatorial plane, to calculate different stellar wind properties and their variation (e.g., \textit{min}, \textit{mean}, \textit{max}) at the distance of interest (see Table \ref{table_3}). In this manner the characterisation preserves the three-dimensional structure of the stellar wind, within the inclination range where the vast majority of exoplanets have been detected to date\footnote[2]{\url{http://exoplanets.org/}} (\citeads{2014PASP..126..827H}).   

The simulated solar wind parameters at 1 AU are consistent with the nominal measurements performed by the \textit{Advanced Composition Explorer (ACE)} spacecraft during different periods of activity. However, the wind particle density ($n$) is too high for the solar maximum simulation\footnote[3]{This is obtained by comparing the simulated average particle density, $n$, with daily ACE measurements during the CR 1962 (Apr-May 2000). See \url{http://www.srl.caltech.edu/ACE/ASC/level2/new/intro.html}}, which is connected to the overestimated mass loss rate for this epoch (see Sect. \ref{sec_alfven}). Still, the simulated parameters are within feasible limits during periods of high activity (which would also involve transient events such as coronal mass ejections, see \citeads{2012LRSP....9....3W}, \citeads{2014ApJ...790...57C}). 

The results listed in Table \ref{table_3} allow a quick assessment of the circumstellar conditions among the different stellar cases and the Sun. For this, one can use the total pressure values associated to the stellar wind in each case ($P_{\rm T}$). This quantity encompasses the thermal properties of the incident plasma ($P_{\rm gas} = nk_{\rm B}T$, with $k_{\rm B}$ as the Boltzmann constant), the dynamic pressure of the wind ($P_{\rm dyn} = nu_{\rm sw}^2/2)$, and the contribution from the magnetic pressure ($P_{\rm mag} = B^2/8\pi$). For instance, if the Earth were located at 1 AU from HD\,1237, the total pressure acting over the magnetosphere would be on average one order of magnitude larger than the nominal conditions around the Sun. In particular sectors of the orbit (i.e., dense streamers in the SH-ZDI simulation), this value can reach more than 2 orders of magnitude of difference compared to the ambient solar wind. This would imply a reduction in the magnetosphere size by a factor $\sim\,2.2$\footnote[4]{This is estimated using Eq. \eqref{eq_2}, replacing the dynamical pressure of the wind ($nu_{\rm sw}^2$) for the total pressure $P_{\rm T}$}. While variations of this order have been observed in the actual magnetosphere of the Earth (see \citeads{2007LRSP....4....1P}), this estimate does not include the effects from magnetic reconnection at the magnetopause (\citeads{2003Natur.426..533F}), or due to transient events such as CMEs (\citeads{2007AsBio...7..167K}; Cohen et al. \citeyearads{2014ApJ...790...57C}, \citeyearads{2011ApJ...738..166C}), which can disrupt the magnetospheric structure. This is extremely important in the case of highly active stars, as these transient events may even dominate completely the mass loss rate, wind, and energetic properties of the environment in these systems (see \citeads{2013ApJ...764..170D}).

\begin{sidewaystable}
\centering
\caption{Stellar wind characterisation at the inner edge of the HZs of the considered systems. The Optimistic (O$_{\rm I-HZ}$) and Conservative (C$_{\rm I-HZ}$) HZ limits have been taken from the HZG (\citeads{2012PASP..124..323K}). Reference calculations at 1 AU are also included. The semi-major axes of Mercury ($a_{\scriptsize\mercury}$) and Venus ($a_{\scriptsize\venus}$) are used in the solar cases. For each reference distance, the parentheses contain the (\textit{min}, \textit{mean}, \textit{max}) values of a given parameter, calculated over a ring extending $\pm\,5^{\circ}$ in inclination from the equatorial plane (see Fig. \ref{fig_12}).}\label{table_3}
\vspace{-5pt}
{\fontsize{8}{10}\selectfont          
\begin{tabular}{ c | c c | c c | c | c c }    
\hline\hline
Parameters &  \multicolumn{2}{|c|}{HD 1237} & \multicolumn{2}{|c|}{HD 22049} & HD 147513 &  \multicolumn{2}{|c}{Sun}\\
 & ZDI & SH-ZDI & ZDI & SH-ZDI & SH-ZDI & CR 1922 (Min) & CR 1962 (Max) \\
\hline
& & & & & & &\\[-8pt]
\textit{Ref. Distance:} & \multicolumn{2}{|c|}{$D =$ 1.0 AU} & \multicolumn{2}{|c|}{$D = $ 1.0 AU} & $D =$ 1.0 AU & \multicolumn{2}{|c}{$D =$ 1.0 AU ($a_{\oplus}$)} \\
$u_{\rm r}$ [km s$^{-1}$] & (586, 983, 1191) & (601, 812, 925) & (522, 808, 1040) & (421, 642, 756) & (259, 433, 670) & (240, 355, 527) & (146, 265, 449) \\
$n$ [cm$^{-3}$] & (3.25, 9.15, 62.9) & (13.7, 33.8, 182.4) & (3.22, 9.48, 48.3) & (15.2, 39.0, 212.0) & (11.5, 51.2, 380.7) & (5.44, 17.9, 125.8) & (15.0, 76.1, 910.0) \\
$T$ [$\times\,10^{4}$ K] & (3.43, 18.1, 29.1)	& (4.05, 12.8, 19.8)	& (3.05, 11.4, 20.1)	& (3.76, 7.24, 11.1) & (0.72, 2.70, 10.5) & (0.79, 1.93, 7.98) & (0.29, 1.48, 13.0) \\
$B$ [nT] & (1.3$\,\times\,10^{-2}$, 6.34, 13.4) & (5.7$\,\times\,10^{-2}$, 17.4, 29.3) & (1.8$\,\times\,10^{-2}$, 4.43, 9.4)	 & (8.8$\,\times\,10^{-2}$, 13.0, 21.0) & (6.6$\,\times\,10^{-2}$, 4.07, 28.4) & (2.5$\,\times\,10^{-3}$, 1.71, 10.8) & (1.6$\,\times\,10^{-2}$, 4.87, 44.3) \\
$P_{\rm T}$ [nPa]\,\tablefootmark{a} & (7.07, 11.2, 38.0) & (17.7, 30.3, 115.9) & (4.58, 8.29, 28.3) & (9.95, 24.4, 139.2) & (4.22, 11.8, 68.7) & (1.44, 3.41, 14.6) & (1.81, 8.56, 84.0) \\
\hline
& & & & & & &\\[-8pt]
\textit{Ref. Distance:} C$_{\rm I-HZ}$ & \multicolumn{2}{|c|}{$D =$ 0.76 AU} & \multicolumn{2}{|c|}{$D = $ 0.58 AU} & $D =$ 0.84 AU & \multicolumn{2}{|c}{$D =$ 0.72 AU ($a_{\venus}$)} \\
$u_{\rm r}$ [km s$^{-1}$] & (582, 991, 1184) & (597, 818, 923) & (517, 7.99, 1032) & (415, 634, 750) & (256, 441, 669) & (229, 356, 531) & (133, 268, 460) \\
$n$ [cm$^{-3}$] & (5.68, 14.3, 97.7) & (23.9, 52.9, 315.0) & (9.86, 28.3, 114.3) & (48.3, 119.4, 427.3) & (16.7, 69.5, 381.1) & (11.8, 34.1, 184.6) & (36.0, 144.2, 1770.0) \\
$T$ [$\times\,10^{4}$ K] & (5.19, 25.8, 39.9)	& (5.97, 18.2, 27.0)	& (7.12, 21.6, 38.0) & (8.07, 13.7, 20.7)	 & (0.92, 3.44, 10.9) & (1.29, 2.97, 8.86) & (0.54, 2.10, 12.0) \\	
$B$ [nT] & (1.7$\,\times\,10^{-2}$, 10.6, 17.8) & (8.5$\,\times\,10^{-2}$, 29.1, 42.9) & (0.10, 12.1, 17.5) & (7.5$\,\times\,10^{-2}$, 35.4, 45.7) & (3.8$\,\times\,10^{-2}$, 5.31, 25.9) & (7.2$\,\times\,10^{-3}$, 2.75, 12.0) & (2,6$\,\times\,10^{-2}$, 7.40, 57.9)\\						
$P_{\rm T}$ [nPa] & (12.6, 18.4, 59.5) & (31.5, 49.1, 196,0) & (16.7, 24.5, 61.2) & (35.8, 73.5, 295.5) & (8.07, 16.5, 75.7) & (3.45, 6.50, 20.9) & (4.53, 15.7, 137.0) \\
\hline
& & & & & & &\\[-8pt]
\textit{Ref. Distance:} O$_{\rm I-HZ}$ & \multicolumn{2}{|c|}{$D =$ 0.60 AU} & \multicolumn{2}{|c|}{$D = $ 0.46 AU} & $D =$ 0.67 AU & \multicolumn{2}{|c}{$D =$ 0.39 AU ($a_{\mercury}$)} \\
$u_{\rm r}$ [km s$^{-1}$] & (579, 960, 1180)	 & (595, 798, 918) & (512, 7.99, 1021) & (412, 634, 746) & (254, 425, 667) & (210, 353, 524) & (116, 269, 467) \\
$n$ [cm$^{-3}$] & (9.37, 26.9, 153.6) & (38.7, 99.0, 504.2) & (15.9, 44.3, 177.4) & (78.9, 182.6, 693.5) & (26.9, 117.8, 464.9) & (42.5, 118.0, 353.4) & (122.0, 515.6, 2614.4) \\
$T$ [$\times\,10^{4}$ K] & (7.36, 32.0, 52.2)	& (8.23, 22.8, 35.0)	& (10.0, 28.7, 49.7)	& (10.6, 18.3, 27.2)	 & (1.34, 4.23, 11.9) & (3.37, 6.53, 11.4) & (1.61, 4.62, 11.1) \\
$B$ [nT] & (3.4$\,\times\,10^{-2}$, 16.4, 24.9) & (7.1$\,\times\,10^{-2}$, 45.6, 62.2) & (7.7$\,\times\,10^{-2}$, 18.9, 25.4) & (0.18, 54.9, 68.4) & (0.15, 7.59, 24.9) & (2.65$\,\times\,10^{-2}$, 7.49, 17.8) & (9.5$\,\times\,10^{-2}$, 18.2, 61.3) \\						
$P_{\rm T}$ [nPa] & (20.5, 31.8, 92.8) & (51.9, 86.0, 309.1) & (26.9, 38.2, 93.2) & (60.4, 112.4, 445.6) & (16.1, 26.0, 81.5) & (16.0, 22.2, 39.5) & (22.1, 53.5, 233.0) \\[1pt]
\hline                  
\end{tabular}} 
\vspace{-5pt}\tablefoot{\tablefoottext{a}{Total pressure of the stellar wind ($P_{\rm T} = P_{\rm gas} + P_{\rm dyn} + P_{\rm mag}$).}} 
\end{sidewaystable}

Finally, this approach can be used to consider the stellar wind properties of the host-star to improve the different estimates of the HZs of these and other systems. One possibility might involve the inclusion into the HZ of a minimum planetary magnetic moment (consistent with the simulated stellar wind conditions), in order to sustain a magnetosphere up to a certain height (e.g. a few planetary radii). This new characteristic of the HZ boundary could even depend on the specific level of coronal activity of the planet-host (which can also be constructed using ZDI-driven models, see \citeads{2016A&A...588A..28A}). This is important due to the fact that in the case of very active planet-hosts (specially M-dwarfs), the magnetic shielding has to compensate the atmospheric expansion induced by the enhanced high-energy emission of the star (e.g. EUV, X-rays, see \citeads{2007AsBio...7..185L}), and stronger CMEs with an increased impact rate (see \citeads{2016arXiv160502683K}). Such dynamic characterisation of the HZ is out of the scope of this paper but will be further explored in a future parametric study, including also additional systems for which ZDI maps are available. 

\vspace{-5pt}
\section{Summary and conclusions}\label{sec_Summary}

\noindent We carried out elaborated simulations of the stellar wind and inner astrospheric structure of three planet-hosting stars (HD\,22049, HD\,1237, and HD\,147513), using the Space Weather Modelling Framework (SWMF, T\'oth et al. \citeyearads{2005JGRA..11012226T}, \citeyearads{2012JCoPh.231..870T}). This paper complements the study presented in \citetads{2016A&A...588A..28A}, which contains the results of the coronal structure modelling of these systems. Steady-state solutions were obtained for two coupled simulation domains, ranging from $1 - 30$ $R_{*}$ (SC domain) and from $25 - 215$ $R_{*}$ (IH domain). Large-scale magnetic field maps of these stars, recovered with Zeeman-Doppler imaging, serve to drive the solutions inside the SC domain, which are coupled self-consistently for a combined solution in the IH domain. A summary of our results and main conclusions is provided below:

\begin{itemize}
\item [-] Following \citetads{2016A&A...588A..28A}, simulations driven by two sets of similar large-scale magnetic field distributions (i.e. ZDI and SH-ZDI) were compared. It is worth noting that both sets of magnetic field maps provided equivalently good fits to the observations and showed substantial similarities in the overall structure of the stellar wind. However, several differences in the magneto-hydrodynamic properties of the solutions were found, including $\sim$10\,--\,30\% denser and $\sim$25\,--\,35\% colder stellar winds in the SH-ZDI solutions compared to the ZDI cases. In addition, the SH-ZDI simulations led to larger values in the average Alfv\'en surface size (by a factor of $\sim$1.5), the mass loss rate $\dot{M}$ (by a factor of $\sim\,$3\,--\,4), and the angular momentum loss rate $\dot{J}$ (by roughly one order of magnitude). Therefore, the values listed in Table \ref{table_1} should be actually interpreted as predicted ranges from this ZDI-driven model.

These variations arise as a consequence of the available magnetic energy to heat the corona and accelerate the wind, which in turn, relates to the different field strengths and map completeness provided by the ZDI and SH-ZDI reconstructions. This strongly differs from previous studies where older implementations of the numerical code used here are considered, and where the completeness in the driving magnetic field distributions yield no significant changes in the wind structure (e.g. \citeads{2012MNRAS.423.3285V}; see Sect. \ref{sec_alfven}). 

\item[-] The results from two different solar simulations, covering activity minimum (CR\,1922) and maximum (CR\,1962), were also considered. We showed that this numerical framework properly recovers the expected structure of the solar wind, including thermodynamical properties (e.g. density, temperature), mass loss and angular momentum loss rates ($\dot{M}_{\odot}$ and $\dot{J}_{\odot}$, respectively; see Table \ref{table_1}), and global topology during each activity state (Figs. \ref{fig_1} and \ref{fig_5}). However, the solar maximum simulation showed an over-enhanced plasma density at 1 AU (Sect. \ref{sec_wind-HZ}, Table \ref{table_3}), as a consequence of an overestimated mass loss rate (by $\sim$\,40\%)\footnote[2]{This corresponds to a very rough estimate, as it relies on the single spatial point measurements and location of Voyager II as reference (\url{http://voyager.jpl.nasa.gov/mission/weekly-reports/index.htm})}. This is interpreted as the result of a considerable fraction of missing mixed polarity regions in the driving magnetogram, which was artificially degraded for a more consistent comparison with the stellar cases (see \citeads{2016A&A...588A..28A}).

\item[-] In general, the stellar wind solutions showed a clear relation with the driving magnetic field distribution, and the developed coronal structure in each case. For HD\,22049 (Figs. \ref{fig_2} and \ref{fig_6}) and HD\,1237 (Figs. \ref{fig_4}, \ref{fig_8} and \ref{fig_9}) various fast-wind regions appeared self-consistently in the simulations, nearly perpendicular to the astrospheric current sheet structure (defined by $B_{\rm r}$ = 0), and with a spatial correspondence with the dominant features in their lower corona (e.g. coronal holes, see \citeads{2016A&A...588A..28A}). The radial wind velocity in these regions reached up to $\sim$\,1100 km s$^{-1}$ in the ZDI simulations, dropping to $\sim$\,700 km s$^{-1}$ in the SH-ZDI cases (see Sect. \ref{sec_resultsIH}). 

\item[-] On the other hand, the simulation of HD\,147513 yielded a much more complex wind solution (Figs. \ref{fig_3} and \ref{fig_7}), compared to what could have been expected from the simple magnetic field distribution driving the simulation (\citeads{2016A&A...588A..28A}). A highly warped astrospheric current sheet was obtained in this case, over which a dominant slow-wind component ($u_{\rm r}$\,$\simeq$\,500 km s$^{-1}$) was developed. While these results could be affected by the comparatively low-resolution of the SH-ZDI map driving the simulation (see \citeads{2016A&A...585A..77H}), this example indicates that numerical descriptions based on first order extrapolations of surface magnetic field properties alone, cannot provide a complete picture of the wind complexity in a given system (e.g. \citeads{2012ApJ...754L..26M}).

\item[-] For HD\,1237 we investigated in detail the wind environment and the conditions experienced by the exoplanet of this system (Sect. \ref{sec_HD1237b}). For this purpose we additionally coupled the Global Magnetosphere (GM) module of the SWMF, to the developed wind structure inside the IH domain. For each simulation (i.e. ZDI and SH-ZDI cases), two representative spatial locations were considered (Figs. \ref{fig_8} and \ref{fig_9}). These included a low-density, fast wind stream, and a high-density, slow wind region. This analysis showed that the density structure of the stellar wind dominates, over the wind velocity, the process of particle injection into the planetary atmosphere (see Table \ref{table_2}). This is a consequence of the large density gradients obtained in the wind solutions (i.e., dense streamers with increments up to 4 orders of magnitude in $n$), compared to the relatively narrow range of resulting radial wind speeds (with variations only up to a factor of 2 in $u_{\rm r}$ for the entire 3D domain).    

\item[-] Following \citeads{2005MNRAS.356.1053S}, we additionally calculated the amount of exoplanetary radio emission from the wind-magnetospheric interaction in this system. We obtained a maximum radio flux on Earth of $F_{\oplus}^{R} \simeq 12$ mJy at 40 MHz, associated with a high-density streamer crossing during periastron passage (at 0.25 AU, \citeads{2001A&A...375..205N}). This value is reduced by an order of magnitude during the orbital motion of the planet (approximately at 2/3 of a right-hand oriented orbit with respect to periastron, see Fig. \ref{fig_10}). Our maximum emission prediction is lower by a factor of $\sim$\,2 compared to the estimates of \citetads{2005MNRAS.356.1053S}, which were based on the mass loss-activity relation of \citeads{2002ApJ...574..412W}, and the assumption of a spherically symmetric wind. Given the system's low declination, SKA is the only facility which could robustly detect and analyse this emission.

\item[-] From our simulations, and applying the methodology explained in \citetads{2014ApJ...783...55C} and \citetads{2015ApJ...813...40G}, we calculated the mass loss rate, $\dot{M}$, and angular momentum loss rate, $\dot{J}$, in these systems. We obtained absolute $\dot{M}$ values, ranging from approximately 1\,$\dot{M}_{\odot}$ up to $\sim$\,7\,$\dot{M}_{\odot}$, and $\dot{J}$ within a broader range of $\sim$\,1--\,60 times the solar prediction (Table \ref{table_1}). In combination with the results for the coronal structure (\citeads{2016A&A...588A..28A}), we constructed, for the first time, a fully simulated mass loss-activity relation, expressed as $\dot{M}_{\rm (sim)} \propto F_{\rm X\,(sim)}^{\rm \,\gamma}$ with $\gamma = 0.79^{\,+\,0.19}_{\,-\,0.15}$. A thoughtful discussion is presented in Sect. \ref{sec_B-Mloss}, comparing this result with the observational relation of \citetads{2005ApJ...628L.143W} (e.g. $\dot{M} \propto F^{1.34 \pm 0.18}_{\rm X}$), exploring various possibilities that could explain the discrepancy in these relations.

\item[-] Finally, by exploiting the 3D capabilities of our simulations we characterised the stellar wind structure at the inner edge of the Habitable Zone (HZ) of these systems (Sect. \ref{sec_wind-HZ}). The optimistic and conservative limits of this boundary, provided in the Habitable Zone Gallery (HZG, \citeads{2012PASP..124..323K}), were considered. We included a 10$^{\circ}$ range in orbital inclination (e.g. Fig. \ref{fig_12}), in order to provide more realistic stellar wind parameters (allowing possible off-the-equator variations), and to capture the region where the majority of exoplanets have been found so far (\citeads{2014PASP..126..827H}). The results of this characterisation are presented in Table \ref{table_3}, and consider all the magneto-hydrodynamic properties of the stellar wind in these systems. Using the solar simulations, reference calculations at the locations of Mercury, Mars, and the Earth are also provided. These results will be used in a future study to perform a dynamical parametrisation of the inner edge of the HZ in these and other systems, accounting for the effects due to the stellar wind and the high-energy environment of the host star.             

\end{itemize}

\begin{acknowledgements}
\noindent We would like to thank the referee for their constructive comments which helped to improve the quality of this paper. This work was carried out using the SWMF/BATSRUS tools developed at The University of Michigan Center for Space Environment Modeling (CSEM) and made available through the NASA Community Coordinated Modeling Center (CCMC). We acknowledge the support by the DFG Cluster of Excellence "Origin and Structure of the Universe". We are grateful for the support by A. Krukau through the Computational Center for Particle and Astrophysics (C2PAP).
\end{acknowledgements}

\vspace{-10pt}
\bibliographystyle{aa}
\bibliography{Biblio}

\end{document}